\documentclass[useAMS,usenatbib,usegraphicx]{mn2e}
\usepackage{longtable}   
\usepackage[dvips]{graphicx}
\usepackage{color}

\title[Identification of SMGs in the SSA22]{
AzTEC/ASTE 1.1 mm survey of SSA22: 
\newline
 Counterpart identification and photometric redshift survey of submillimeter galaxies
}
\author[H. Umehata et al]{H. Umehata$^{1}$\thanks{E-mail:
umehata@ioa.s.u-tokyo.ac.jp}
Y. Tamura$^{1}$,
K. Kohno$^{1,13}$,
B. Hatsukade$^{2}$,
K. S. Scott$^{3}$,
M. Kubo$^{4}$,
\newauthor
T. Yamada$^{4}$,
R. J. Ivison$^{5,6}$ 
R. Cybulski$^{7}$,
I. Aretxaga$^{8}$,
J. Austermann$^{12}$,
\newauthor
D. H. Hughes$^{8}$,
H. Ezawa$^{2}$,
T. Hayashino$^{4}$
S. Ikarashi$^{1}$,
D. Iono$^{2,9}$,
\newauthor
R. Kawabe$^{2,10}$,
Y. Matsuda$^{2}$,
H. Matsuo$^{14}$
K. Nakanishi$^{2,10,11}$,
T. Oshima$^{9}$,
\newauthor
T. Perera$^{7}$,
T. Takata$^{2}$,
G. W. Wilson$^{7}$, and
M. S. Yun$^{7}$
\\
$^{1}$Institute of Astronomy, The University of Tokyo, Mitaka, Tokyo 181-0015, Japan\\
$^{2}$National Astronomical Observatory of Japan, Mitaka, Tokyo 181-8588, Japan\\
$^{3}$North American ALMA Science Center, National Radio Astronomy Observatory, 520 Edgemont Road, Charlottesville, Virginia,\\
 22903, USA\\
$^{4}$Astronomical Institute, Tohoku University, 6-3 Aoba, Aramaki, Aoba-ku, Sendai, Miyagi, 980-8578, Japan\\
$^{5}$UK Astronomy Technology Centre, Science and Technology Facilities Council, Royal Observatory, Blackford Hill, \\
Edinburgh EH9 3HJ \\
$^{6}$Institute for Astronomy, University of Edinburgh, Blackford Hill, Edinburgh EH9 3HJ\\
$^{7}$Department of Astronomy, University of Massachusetts, Amherst, MA 01003, USA\\
$^{8}$Instituto Nacional de Astrofisica, Optica y Electronica (INAOE), Aptdo. Postal 51 y 216, 72000 Puebla, Pue., Mexico\\
$^{9}$Nobeyama Radio Observatory, National Astronomical Observatory of Japan, Minaminaki, Minamisaku, Nagano 384-1305, Japan\\
$^{10}$Joint ALMA Observatory, Alonso de Cordova 3107, Vitacura, Santiago 763 0355, Chile\\
$^{11}$The Graduate University for Advanced Studies (Sokendai), Mitaka, Tokyo 181-8588, Japan\\
$^{12}$Center for Astrophysics and Space Astronomy, University of Colorado\\
$^{13}$Research Center for the Early Universe (WPI), University of Tokyo, 7-3-1 Hongo, Bunkyo, Tokyo 113-0033, Japan\\
$^{14}$Advanced Technology Center, National Astronomical Observatory of Japan, Mitaka, Tokyo 181-8588,
Japan
}
\begin{document}

\date{Accepted 2014?}

\pagerange{\pageref{firstpage}--\pageref{lastpage}} \pubyear{2012}

\maketitle

\label{firstpage}

\begin{abstract}

We present the results from a 1.1~mm imaging survey of the SSA22 field,
known for having an overdensity of $z=3.1$ Lyman-alpha emitting galaxies (LAEs), 
taken with the AzTEC camera on the Atacama Submillimeter Telescope Experiment (ASTE).
We imaged a 950 arcmin$^2$ field down to a 1$\sigma$ sensitivity of 0.7--1.3 mJy/beam to find 125 submillimeter galaxies (SMGs) with a signal to noise ratio $\ge$ 3.5.
%In addition, 17 SMGs, located just outside of the region, were also listed as supplementary sources. 
Counterpart identification using radio and near/mid-infrared data was performed and 
one or more counterpart candidates were found for 59 SMGs.
Photometric redshifts based on optical to near-infrared images were evaluated for 45 SMGs of these SMGs with {\it Spitzer}/IRAC data, and the median value is found to be $z=2.4$.
By combining these estimation with estimates from the literature we determined that 10 SMGs might lie within the large-scale structure at $z=3.1$.
The two-point angular cross-correlation function between LAEs and SMGs indicates that the positions of the SMGs are correlated with the $z=3.1$ protocluster.
These results suggest that the SMGs were formed and evolved selectively in the high dense environment of the high redshift universe. This picture is consistent with the predictions of the standard model of hierarchical structure formation. 
\end{abstract}


\section{Introduction}
The first deep extragalactic survey at 850 $\mu$m undertaken with the Submillimeter Common-User Bolometer Array (SCUBA, \citealt{1999MNRAS.303..659H}) on the James Clerk Maxwell Telescope (JCMT) unveiled a population of galaxies in the distant redshift universe that are extremely bright at submillimeter wavelengths (\citealt{1997ApJ...490L...5S}, \citealt{1998Natur.394..241H}, \citealt{1998Natur.394..248B}).
This was followed by several wide surveys at (sub)millimeter wavelengths have been conducted to discover more and more such galaxies (e.g., \citealt{2004MNRAS.354..779G}, \citealt{2008MNRAS.385.2225S}, \citealt{2009MNRAS.395.1905C}, \citealt{2011MNRAS.411..102H}).

These submillimeter bright galaxies (SMGs) have huge rest-frame far-infrared (FIR) luminosities ($L_{\mathrm{FIR}}\sim10^{12}-10^{13}L_{\odot}$), which should be caused mainly by highly dust-enshrouded star formation and are  often indicative of a star-formation rate (SFR) of $\geq$ 1000 M$_\odot$/yr.
Their flux at submillimeter wavelengths is almost constant for galaxies with a given FIR luminosity at $z\sim1$--10 due to the negative $k$-correction. It is hence of great benefit to discover high redshift objects (for review, see \citealt{2002PhR...369..111B}). 
In addition to their high activity,
spectroscopic observations of the millimeter and submillimeter transitions of molecular carbon monoxide (CO) have unveiled the SMGs' large dynamical and gas masses (e.g., \citealt{2005MNRAS.359.1165G}).
This observational evidence shows that SMGs were the most active, massive star forming galaxies in the early universe.
The clarification of the SMG nature and formation process is seriously important to understand the galaxy formation and evolution.

As with other galaxy populations, one of the most crucial questions about the SMGs is how their formation and evolution depend on their environment. 
The current cold dark matter (CDM) cosmological simulations show that the SMGs 
should preferentially exist in regions where the mass densities are high and, correspondingly, also the merger rates are high (e.g., \citealt{2005Natur.435..629S}).
SMGs are also supposed to be progenitors of the massive elliptical galaxies seen in the cores of present-day rich clusters (e.g., \citealt{1999ApJ...515..518E}).
While the connection between SMGs and massive dark matter haloes has been statistically indicated by clustering analysis (e.g., \citealt{2004ApJ...611..725B}; \citealt{2012MNRAS.421..284H}),
the connection in individual cluster/protocluster is still unclear, although previous papers have reported some cases.
\cite{2011Natur.470..233C} and \cite{2009ApJ...694.1517D} reported the discovery of SMGs in overdense region at $z=5.3$ and $z=4.05$, respectively. 
These results indicate that the overdense regions might be sites of SMG formation.
On the other hand, \cite{2009ApJ...691..560C} shows that SMGs are formed in less overdence regions at $z=1.99$.
Thus the environmental dependence on SMG formation is still controversial.
%%NOW%%

The relationship between SMGs and the surrounding environment is also intriguing from the view point of galaxy formation at (proto)clusters.
In the local universe the morphology-density relation has been well-known.
Observations have revealed a higher fraction of early-type galaxies in denser environments \citep{1980ApJ...236..351D}.
This trend has been confirmed for up to $z\sim$1 universe (e.g., \citealt{2005ApJ...623..721P} ).
 Although this is one of the most established environmental effects on galaxy evolution, it is difficult to examine the relation directly at higher redshifts.
Instead, the color-density or color-magnitude relations were examined as proxies.
For instance, \cite{2007MNRAS.377.1717K} examined the color-magnitude relation in protoclusters and found that the red sequence of galaxies, which is well-established in clusters at least out to $z\sim1$, first appeared at $z=2-3$. This suggests that 
massive galaxies were assembled in protoclusters in this era.

SSA22 is a unique laboratory field to investigate the formation of star-forming galaxies including SMGs in the overdense region.
\cite{1998ApJ...492..428S} first discovered this protocluster as a concentration of Lyman break galaxies (LBGs) at $z = 3.09$.
Furthermore \cite{2000ApJ...532..170S} found that the surface density of Lyman alpha emitters (LAEs) was also much higher than that of other fields.
Consequently, the wide field survey using Subaru/Suprime-Cam equipped with a narrowband filter (NB497) have revealed a structure  which was traced by LAEs and spread over 700 arcmin$^2$(\citealt{2004AJ....128.2073H}; \citealt{2012AJ....143...79Y}).
\cite{2012AJ....143...79Y} evaluated that the degree of overdensity was at most 10 times of the expected standard deviations based on the counts of LAEs at $z=3.1$.
Hence the SSA22 field can provide us with unique insights regarding galaxy formation in an overdense environment. 
Several previous works on SMGs discovered by SCUBA and AzTEC/ASTE surveys in this field has been reported. 
\cite{2004ApJ...611..725B} and \cite{2005ApJ...622..772C} confirmed three SCUBA SMGs have $z_{spec}=3.1$ and these SMGs really lie within the densest region. 
\cite{2009Natur.459...61T} showed that there was an angular correlation between the 15 brightest AzTEC SMGs and $z=3.1$ LAEs.

But the lack of redshift information prevents us from investigating this further.
Although redshift is one of the essential pieces of information for this purpose, 
obtaining them has remained a difficult task. Firstly, the typical beam size of the single-dish telescopes used for wide field SMG surveys is insufficient to determine position accurately.
In the case of the AzTEC/ASTE survey, we can achieve only $\sim30^{\prime\prime}$ FWHM.
Accurate positions of SMGs could ideally be obtained with a sub-millimeter interferometer, like the Atacama Large Millimeter/submillimeter Array (ALMA; \citealt{2010SPIE.7733E..34H}), but such observations are time-intensive over large fields.
Secondly SMGs generally tend to be optically faint due to dust attenuation and hence 
optical observations are often helpless to determine counterpart and obtain redshift information.
To overcome such difficulties in searching counterparts and determining redshift, 
previous works have shown that 
multi-wavelength identification utilizing radio, MIPS(\citealt{2004ApJS..154...25R}), and IRAC(\citealt{2004ApJS..154...10F}) imaging data is useful (e.g., {\citealt{2011MNRAS.413.2314B}; \citealt{2011MNRAS.415.1479W}; \citealt{2012MNRAS.420..957Y}).
If the identified counterparts in these images have an optical to near-infrared counterpart, we can derive a photometric redshift. This is the approach we will follow. We expand the area concerned by Tamura et al. and add photometric redshift information to investigate the relation between SMGs and underlying environment more closely.

The organization of this paper is as follows. In Section 2 we report the general properties of the AzTEC/ASTE survey and the detected SMGs. The utilization of other wavelength data set of this field is presented in Section 3. In Section 4 we describe our analysis used for counterpart identification.
The estimation of the photometric redshifts are derived in Section 5.
In Section 6 we discuss the relationship between SMGs and the $z=3.1$ protocluster.
We assume a cosmology with $\Omega_m=0.3, \Omega_\lambda=0.7, H_0=70$ km s$^{-1}$ Mpc$^{-1}$ and all magnitudes are given according to the AB system throughout the paper.

\section{AzTEC/ASTE Observation}
\subsection{Observations}
%% - Instrument
We used the AzTEC camera operating at 1.1 mm \cite{2008MNRAS.386..807W}  mounted on the ASTE 10-m submillimetre telescope (\citealt{2004SPIE.5489..763E}, \citealt{2008SPIE.7012E...6E}) located on Pampa la Bola, near Cerro Chajnantor in northern Chile.  The AzTEC camera is a 144-element bolometer array and AzTEC/ASTE provides an angular resolution of $28''$ in full width at half maximum (FWHM).
%% - Runs
All of the AzTEC/ASTE observations of SSA22 were carried out at night in August--September 2007 and August--September 2008. Part of these data were reported in
 \cite{2009Natur.459...61T} and \cite{2013MNRAS.430.2768T}.  The dry weather at the high site (4860 m in elevation) provided an excellent data set that was taken under $\tau_{\rm 220\,GHz} = 0.01$--0.10.
%% - Scan patterns
We mapped a 50-arcmin diameter region centered at R.A.\ (J2000) = 22h 17.6m, Decl.\ (J2000) = +0$^{\circ}$ 15.0$'$. Observations were made using the on-the-fly (OTF) mode by continuously scanning the telescope boresight in azimuth and elevation in a modified Lissajous scan pattern, which is described as a function of time $t$ by,
\begin{eqnarray}
\delta Az &=& A \sin{at} + B \sin(at / f)\\
\delta El &=& C \sin{bt} + D \sin(bt / f),
\end{eqnarray}
In 2007, we use a scan pattern with $(A, B, C, D, a, b, f) = (7'40'', 7'40'', 3'40'', 3'40'', 5.0, 4.0, 45.0)$.  The actual values of $a$ and $b$ are normalized to limit the peak telescope slew velocity to 330 arcsec s$^{-1}$. In order to go deeper and wider, we used another scan pattern for the 2008 observations with $(A, B, C, D, a, b, f) = (22', 22', 2', 2', 6.0, 5.0, 23)$.  
The coverage was centered at the same position as those in 2007, but swept a doughnut-like region surrounding the 2007 coverage. The total integrated time on source through 2007--2008 was 74 hr. The area of the co-added map where the noise levels are below 1 mJy was estimated to reach 900 arcmin$^2$.
%% - pointing calibration
The wind speeds during the runs were 1--10 m/s and typically 5 m/s, providing a random telescope pointing offset of $< 1''$.  The astrometry was checked using a bright quasar 3C446, 5.6 deg away from SSA22, every 1-1.5 hr during the observations.  The correction to the telescope pointing model was small (typically $< 2''$) and applied to all scans toward SSA22, resulting in an absolute pointing accuracy better than $4''$ (\citealt{2013MNRAS.430.2768T}).
%% - flux calibration
For flux calibration, beam shape measurements and array flat-fielding, beammaps of Uranus and Neptune were taken at least once, typically twice, a night.  
The beammaps were made so that each bolometer pixel raster-scans and images the planets.  The flux calibration accuracy was estimated to be 10\% or better.

\subsection{Data Reduction}
%% 2.2. Data reduction

%% - reduction
The data were reduced in the standard manner described in \cite{2008MNRAS.385.2225S} and \cite{2012MNRAS.423..529D} , and 
we mention only the outline here. We employed an atmospheric noise removal technique based on principal component analysis (PCA cleaning) to isolate the low-frequency atmospheric noise from the astronomical signals involved in time-stream data. The cleaned time-stream data of each scan were projected into a map by binning them into $3'' \times 3''$ pixels and the individual scans were coadded into a single map by weighted-averaging.
%% - noise map

We also generated 100 noise realizations by jackknifing the time-stream data (i.e., by multiplying each 15-s time-stream interval by $\pm 1$ randomly and then reducing in the standard manner), which were free of astronomical emission and hence represented the underlying photon noises from the atmosphere and instruments.  These random maps were used to estimate a ``noise map'' by computing pixel-to-pixel standard deviations, which represents the local noise level.  We also create a weight map by computing an inverse-square of the noise map.
%% "weight map" は、XX% coverage というところでしか、表面上は用いていないので、どこまで書くべきかちょっと図りかねますが、とりあえず。
%% - PRF
Since the PCA cleaning was AC-coupled to the time-stream and hence worked as a high-pass filter (i.e., the resulting map is zero-mean), it attenuated the peak flux and caused negative sidelobes around a source. To correct these effects, we simulated the profile of the point-source response function (a point source kernel) following the method presented in \cite{2012MNRAS.423..529D}.

\subsection{Map and Source Catalog}

\begin{table}
\begin{center}
\footnotesize
\begin{tabular}{cccccc} 
\hline
\hline
Coverage & Area & Noise level & N(Source)  & N(False) \\
 & [arcmin$^2$] & [mJy beam$^{-1}$] & & \\
\hline
50\% & 749 & 0.72--1.00 & 107 & 6.2$\pm$2.4 \\
30-50\% & 205 &1.00--1.32 & 18 & 2.4$\pm$0.8 \\
\hline
\end{tabular}
\end{center}
\label{map_prop}
\caption{
Map properties in the 50\% and the 30-50\% coverage regions.
N(source) and N(False) show number of sources and false detections ($\leq3.5\sigma$ for both).
}
\end{table}

  \begin{figure*}
    \centering
    \includegraphics[width=1.99\columnwidth]{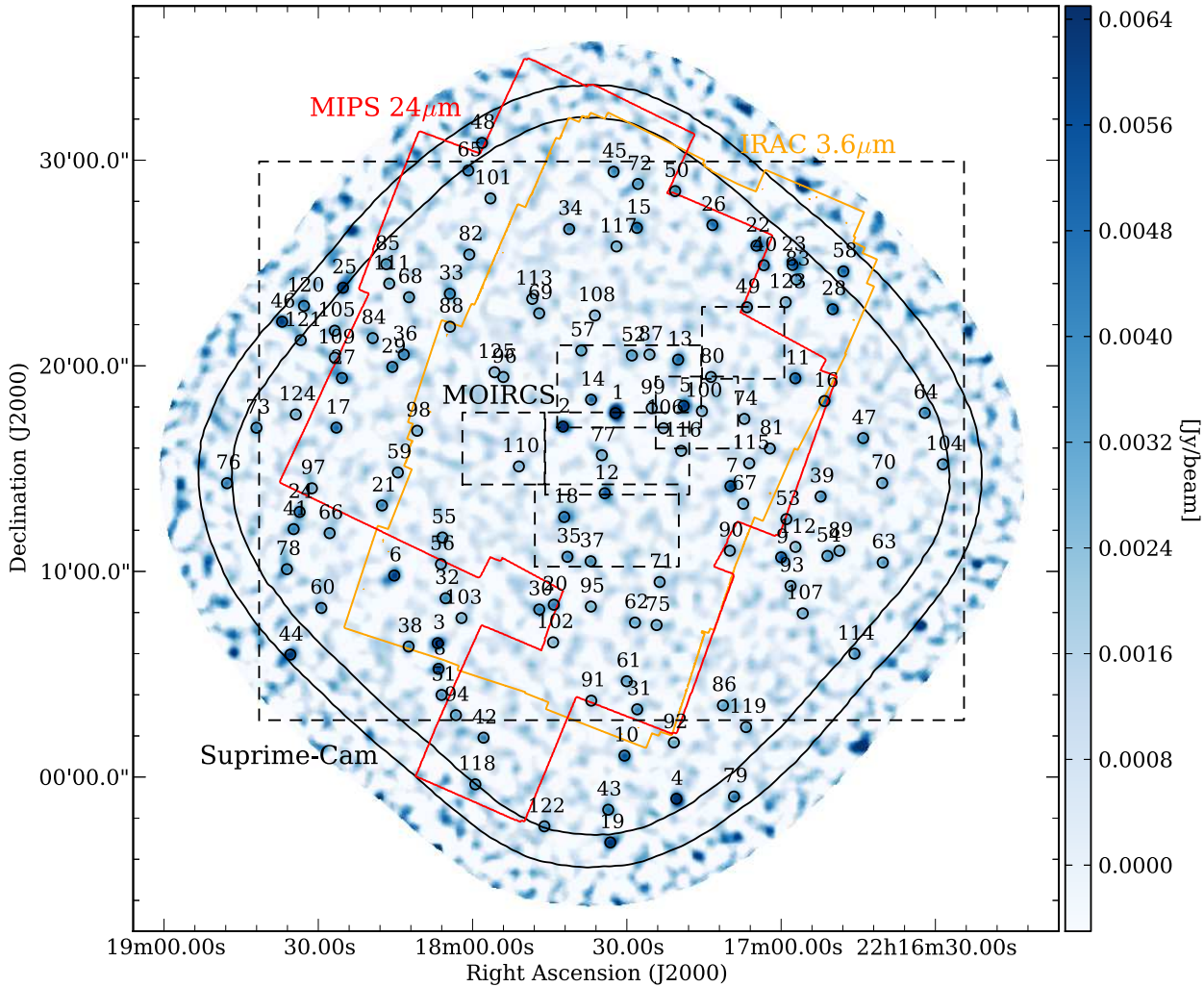}
    \caption{AzTEC/ASTE 1.1 mm image and observation areas of each instrument. The background image is the AzTEC 1.1 mm signal map of the SSA22 field with a 10\% uniform coverage region.
    The side bar represents flux density in unit of Jy/beam.
     The 30\% and 50\% coverage regions are shown using black contours.
      The black circles indicate the positions of $\ge$ 3.5 $\sigma$ sources and the diameter is 30 $^{\prime \prime}$, which correspond to the FWHM of AzTEC/ASTE.
   In addition, the area of Suprime-Cam(large dashed rectangle), MIPS 24 $\mu$m(red line), IRAC 3.6 $\mu$m(yellow line), and MOIRCS(small dashed rectangle) are shown. UKIRT/DXS survey  and VLA 1.4 GHz observations contains all AzTEC sources.}
    \label{aztec_map}
  \end{figure*}
  
\begin{table*}
\begin{center}
\caption{The AzTEC/ASTE SSA22 source catalog.The machine-readable version is available online. The columns give: 1) source name; 2) AzTEC source ID; 3) Right ascension; 4) Declination; 
5) observed 1.1 mm flux density and 1$\sigma$ error; 6) deboosted 1.1 mm flux density and 68\% confidence interval; and 7) signal to noise ratio of the detection in the AzTEC map. }
\begin{tabular}{ccccccc}
\hline
\hline
Name & ID & R.A. & Dec. & S$_{\rm observed}$ & S$_{\rm deboosted}$ & S/N \\ 
&  & [$^{\rm h}$ $^{\rm m}$ $^{\rm s}$ ] & [$^{\circ}$ $^{\prime}$ $^{\prime\prime}$ ] & [mJy] & [mJy] & \\
 \hline
AzTEC J221732.21+001742.1 & SSA22-AzTEC1 & 22 17 32.21    &  +00 17 42.1    & 11.9   $\pm$ 0.7    & 11.3  $^{+  0.9   }_{- 0.7  }$  & 16.2 \\ 
AzTEC J221742.42+001702.5 & SSA22-AzTEC2 & 22 17 42.42    &  +00 17 02.5    & 7.5   $\pm$ 0.7    & 6.9  $^{+  0.9   }_{- 0.7  }$  & 10.1 \\ 
AzTEC J221806.78+000630.6 & SSA22-AzTEC3 & 22 18 06.78    &  +00 06 30.6    & 7.4   $\pm$ 0.8    & 6.9  $^{+  0.8   }_{- 0.9  }$  & 9.4 \\ 
AzTEC J221720.36$-$000103.6 & SSA22-AzTEC4 & 22 17 20.36    &  $-$00 01 03.6    & 9.4   $\pm$ 1.0    & 8.5  $^{+  1.1   }_{- 1.1  }$  & 9.2 \\ 
AzTEC J221718.95+001803.0 & SSA22-AzTEC5 & 22 17 18.95    &  +00 18 03.0    & 6.6   $\pm$ 0.7    & 6.1  $^{+  0.8   }_{- 0.8  }$  & 8.9 \\ 
AzTEC J221815.23+000948.0 & SSA22-AzTEC6 & 22 18 15.23    &  +00 09 48.0    & 6.5   $\pm$ 0.8    & 5.9  $^{+  0.8   }_{- 0.8  }$  & 8.5 \\ 
AzTEC J221709.85+001408.9 & SSA22-AzTEC7 & 22 17 09.85    &  +00 14 08.9    & 6.1   $\pm$ 0.7    & 5.6  $^{+  0.8   }_{- 0.8  }$  & 8.3 \\ 
AzTEC J221806.62+000515.7 & SSA22-AzTEC8 & 22 18 06.62    &  +00 05 15.7    & 6.2   $\pm$ 0.8    & 5.7  $^{+  0.8   }_{- 0.9  }$  & 7.7 \\ 
AzTEC J221700.00+001041.2 & SSA22-AzTEC9 & 22 17 00.00    &  +00 10 41.2    & 5.8   $\pm$ 0.8    & 5.3  $^{+  0.8   }_{- 0.8  }$  & 7.7 \\ 
AzTEC J221730.46+000102.5 & SSA22-AzTEC10 & 22 17 30.46    &  +00 01 02.5    & 5.9   $\pm$ 0.8    & 5.3  $^{+  0.9   }_{- 0.9  }$  & 7.3 \\ 
AzTEC J221657.24+001923.6 & SSA22-AzTEC11 & 22 16 57.24    &  +00 19 23.6    & 5.5   $\pm$ 0.8    & 4.9  $^{+  0.9   }_{- 0.8  }$  & 7.2 \\ 
AzTEC J221734.30+001348.2 & SSA22-AzTEC12 & 22 17 34.30    &  +00 13 48.2    & 5.4   $\pm$ 0.8    & 4.9  $^{+  0.8   }_{- 0.8  }$  & 7.1 \\ 
AzTEC J221720.03+002017.9 & SSA22-AzTEC13 & 22 17 20.03    &  +00 20 17.9    & 5.2   $\pm$ 0.7    & 4.6  $^{+  0.8   }_{- 0.8  }$  & 7.0 \\ 
AzTEC J221736.96+001821.3 & SSA22-AzTEC14 & 22 17 36.96    &  +00 18 21.3    & 5.0   $\pm$ 0.7    & 4.5  $^{+  0.8   }_{- 0.8  }$  & 6.9 \\ 
AzTEC J221728.04+002642.6 & SSA22-AzTEC15 & 22 17 28.04    &  +00 26 42.6    & 5.3   $\pm$ 0.8    & 4.7  $^{+  0.9   }_{- 0.8  }$  & 6.8 \\ 
AzTEC J221651.56+001817.2 & SSA22-AzTEC16 & 22 16 51.56    &  +00 18 17.2    & 5.2   $\pm$ 0.8    & 4.6  $^{+  0.9   }_{- 0.8  }$  & 6.7 \\ 
AzTEC J221826.48+001659.6 & SSA22-AzTEC17 & 22 18 26.48    &  +00 16 59.6    & 5.1   $\pm$ 0.8    & 4.5  $^{+  0.8   }_{- 0.8  }$  & 6.5 \\ 
AzTEC J221742.18+001238.8 & SSA22-AzTEC18 & 22 17 42.18    &  +00 12 38.8    & 4.8   $\pm$ 0.7    & 4.3  $^{+  0.8   }_{- 0.8  }$  & 6.5 \\ 
AzTEC J221733.21$-$000310.8 & SSA22-AzTEC19 & 22 17 33.21    &  $-$00 03 10.8    & 7.5   $\pm$ 1.2    & 6.3  $^{+  1.2   }_{- 1.2  }$  & 6.5 \\ 
AzTEC J221744.20+000822.9 & SSA22-AzTEC20 & 22 17 44.20    &  +00 08 22.9    & 4.5   $\pm$ 0.7    & 4.0  $^{+  0.8   }_{- 0.8  }$  & 6.2 \\ 
AzTEC J221817.62+001312.6 & SSA22-AzTEC21 & 22 18 17.62    &  +00 13 12.6    & 4.6   $\pm$ 0.7    & 4.0  $^{+  0.8   }_{- 0.8  }$  & 6.1 \\ 
AzTEC J221704.79+002550.6 & SSA22-AzTEC22 & 22 17 04.79    &  +00 25 50.6    & 5.7   $\pm$ 0.9    & 4.8  $^{+  1.0   }_{- 1.0  }$  & 6.1 \\ 
AzTEC J221657.77+002454.3 & SSA22-AzTEC23 & 22 16 57.77    &  +00 24 54.3    & 5.9   $\pm$ 1.0    & 5.0  $^{+  1.0   }_{- 1.1  }$  & 6.1 \\ 
AzTEC J221833.64+001253.6 & SSA22-AzTEC24 & 22 18 33.64    &  +00 12 53.6    & 5.0   $\pm$ 0.8    & 4.4  $^{+  0.9   }_{- 0.9  }$  & 6.1 \\ 
AzTEC J221825.20+002347.9 & SSA22-AzTEC25 & 22 18 25.20    &  +00 23 47.9    & 6.6   $\pm$ 1.1    & 5.5  $^{+  1.2   }_{- 1.2  }$  & 6.0 \\ 
AzTEC J221713.37+002650.8 & SSA22-AzTEC26 & 22 17 13.37    &  +00 26 50.8    & 5.1   $\pm$ 0.9    & 4.4  $^{+  0.9   }_{- 1.0  }$  & 5.9 \\ 
AzTEC J221825.40+001924.5 & SSA22-AzTEC27 & 22 18 25.40    &  +00 19 24.5    & 4.9   $\pm$ 0.8    & 4.2  $^{+  0.9   }_{- 0.9  }$  & 5.9 \\ 
AzTEC J221649.97+002245.2 & SSA22-AzTEC28 & 22 16 49.97    &  +00 22 45.2    & 5.5   $\pm$ 0.9    & 4.7  $^{+  1.0   }_{- 1.1  }$  & 5.8 \\ 
AzTEC J221815.60+001956.7 & SSA22-AzTEC29 & 22 18 15.60    &  +00 19 56.7    & 4.5   $\pm$ 0.8    & 3.8  $^{+  0.9   }_{- 0.8  }$  & 5.8 \\ 
AzTEC J221747.03+000809.0 & SSA22-AzTEC30 & 22 17 47.03    &  +00 08 09.0    & 4.2   $\pm$ 0.7    & 3.7  $^{+  0.8   }_{- 0.8  }$  & 5.8 \\ 
AzTEC J221727.99+000317.6 & SSA22-AzTEC31 & 22 17 27.99    &  +00 03 17.6    & 4.3   $\pm$ 0.8    & 3.7  $^{+  0.8   }_{- 0.8  }$  & 5.7 \\ 
AzTEC J221805.25+000841.8 & SSA22-AzTEC32 & 22 18 05.25    &  +00 08 41.8    & 4.2   $\pm$ 0.8    & 3.7  $^{+  0.8   }_{- 0.9  }$  & 5.6 \\ 
AzTEC J221804.42+002330.4 & SSA22-AzTEC33 & 22 18 04.42    &  +00 23 30.4    & 4.4   $\pm$ 0.8    & 3.8  $^{+  0.9   }_{- 0.9  }$  & 5.6 \\ 
AzTEC J221741.21+002639.0 & SSA22-AzTEC34 & 22 17 41.21    &  +00 26 39.0    & 4.3   $\pm$ 0.8    & 3.7  $^{+  0.8   }_{- 0.8  }$  & 5.6 \\ 
AzTEC J221741.56+001042.7 & SSA22-AzTEC35 & 22 17 41.56    &  +00 10 42.7    & 4.1   $\pm$ 0.7    & 3.5  $^{+  0.8   }_{- 0.8  }$  & 5.5 \\ 
AzTEC J221813.38+002032.9 & SSA22-AzTEC36 & 22 18 13.38    &  +00 20 32.9    & 4.2   $\pm$ 0.8    & 3.6  $^{+  0.8   }_{- 0.9  }$  & 5.4 \\ 
AzTEC J221737.05+001029.7 & SSA22-AzTEC37 & 22 17 37.05    &  +00 10 29.7    & 4.0   $\pm$ 0.7    & 3.4  $^{+  0.8   }_{- 0.8  }$  & 5.4 \\ 
AzTEC J221812.44+000620.8 & SSA22-AzTEC38 & 22 18 12.44    &  +00 06 20.8    & 4.3   $\pm$ 0.8    & 3.6  $^{+  0.9   }_{- 0.9  }$  & 5.3 \\ 
AzTEC J221652.32+001338.6 & SSA22-AzTEC39 & 22 16 52.32    &  +00 13 38.6    & 3.9   $\pm$ 0.8    & 3.3  $^{+  0.8   }_{- 0.8  }$  & 5.2 \\ 
AzTEC J221703.36+002453.6 & SSA22-AzTEC40 & 22 17 03.36    &  +00 24 53.6    & 4.6   $\pm$ 0.9    & 3.8  $^{+  1.0   }_{- 1.0  }$  & 5.2 \\ 
AzTEC J221834.80+001203.5 & SSA22-AzTEC41 & 22 18 34.80    &  +00 12 03.5    & 4.5   $\pm$ 0.9    & 3.7  $^{+  0.9   }_{- 1.0  }$  & 5.2 \\ 
AzTEC J221757.87+000154.9 & SSA22-AzTEC42 & 22 17 57.87    &  +00 01 54.9    & 4.5   $\pm$ 0.9    & 3.7  $^{+  0.9   }_{- 1.0  }$  & 5.1 \\ 
AzTEC J221733.63$-$000135.4 & SSA22-AzTEC43 & 22 17 33.63    &  $-$00 01 35.4    & 4.8   $\pm$ 1.0    & 3.9  $^{+  1.0   }_{- 1.0  }$  & 5.1 \\ 
AzTEC J221835.43+000557.3 & SSA22-AzTEC44 & 22 18 35.43    &  +00 05 57.3    & 6.8   $\pm$ 1.3    & 5.0  $^{+  1.5   }_{- 1.4  }$  & 5.0 \\ 
AzTEC J221732.59+002926.4 & SSA22-AzTEC45 & 22 17 32.59    &  +00 29 26.4    & 4.3   $\pm$ 0.9    & 3.5  $^{+  1.0   }_{- 0.9  }$  & 5.0 \\ 
AzTEC J221837.04+002208.7 & SSA22-AzTEC46 & 22 18 37.04    &  +00 22 08.7    & 6.5   $\pm$ 1.3    & 4.8  $^{+  1.5   }_{- 1.4  }$  & 5.0 \\ 
AzTEC J221644.05+001629.1 & SSA22-AzTEC47 & 22 16 44.05    &  +00 16 29.1    & 3.9   $\pm$ 0.8    & 3.2  $^{+  0.9   }_{- 0.9  }$  & 4.8 \\ 
AzTEC J221758.17+003050.5 & SSA22-AzTEC48 & 22 17 58.17    &  +00 30 50.5    & 5.7   $\pm$ 1.2    & 4.3  $^{+  1.3   }_{- 1.3  }$  & 4.8 \\ 
AzTEC J221706.64+002250.9 & SSA22-AzTEC49 & 22 17 06.64    &  +00 22 50.9    & 3.8   $\pm$ 0.8    & 3.1  $^{+  0.9   }_{- 0.9  }$  & 4.8 \\ 
AzTEC J221720.60+002829.8 & SSA22-AzTEC50 & 22 17 20.60    &  +00 28 29.8    & 4.3   $\pm$ 0.9    & 3.4  $^{+  1.0   }_{- 1.0  }$  & 4.8 \\ 
AzTEC J221806.03+000359.6 & SSA22-AzTEC51 & 22 18 06.03    &  +00 03 59.6    & 4.1   $\pm$ 0.9    & 3.3  $^{+  0.9   }_{- 1.0  }$  & 4.7 \\ 
AzTEC J221729.02+002030.2 & SSA22-AzTEC52 & 22 17 29.02    &  +00 20 30.2    & 3.5   $\pm$ 0.7    & 2.9  $^{+  0.8   }_{- 0.8  }$  & 4.7 \\ 
AzTEC J221658.98+001232.7 & SSA22-AzTEC53 & 22 16 58.98    &  +00 12 32.7    & 3.5   $\pm$ 0.7    & 2.8  $^{+  0.9   }_{- 0.8  }$  & 4.7 \\ 
AzTEC J221650.99+001045.2 & SSA22-AzTEC54 & 22 16 50.99    &  +00 10 45.2    & 3.7   $\pm$ 0.8    & 3.0  $^{+  0.9   }_{- 0.9  }$  & 4.6 \\ 
AzTEC J221805.86+001139.0 & SSA22-AzTEC55 & 22 18 05.86    &  +00 11 39.0    & 3.3   $\pm$ 0.7    & 2.7  $^{+  0.8   }_{- 0.8  }$  & 4.6 \\ 
\hline
\end{tabular}
\end{center}
\end{table*}

\setcounter{table}{1}
\begin{table*}
\begin{center}
\caption{-- continued}
\begin{tabular}{ccccccc}
\hline
\hline
Name & ID & R.A. & Dec. & S$_{\rm observed}$ & S$_{\rm deboosted}$ & S/N \\ 
&  & [$^{\rm h}$ $^{\rm m}$ $^{\rm s}$ ] & [$^{\circ}$ $^{\prime}$ $^{\prime\prime}$ ] & [mJy] & [mJy] & \\
 \hline
AzTEC J221806.15+001021.2 & SSA22-AzTEC56 & 22 18 06.15    &  +00 10 21.2    & 3.3   $\pm$ 0.7    & 2.7  $^{+  0.8   }_{- 0.8  }$  & 4.5 \\ 
AzTEC J221738.86+002044.8 & SSA22-AzTEC57 & 22 17 38.86    &  +00 20 44.8    & 3.3   $\pm$ 0.7    & 2.7  $^{+  0.8   }_{- 0.8  }$  & 4.5 \\ 
AzTEC J221647.90+002435.6 & SSA22-AzTEC58 & 22 16 47.90    &  +00 24 35.6    & 5.2   $\pm$ 1.2    & 3.8  $^{+  1.3   }_{- 1.4  }$  & 4.4 \\ 
AzTEC J221814.53+001448.8 & SSA22-AzTEC59 & 22 18 14.53    &  +00 14 48.8    & 3.2   $\pm$ 0.7    & 2.6  $^{+  0.8   }_{- 0.8  }$  & 4.4 \\ 
AzTEC J221829.42+000813.4 & SSA22-AzTEC60 & 22 18 29.42    &  +00 08 13.4    & 4.1   $\pm$ 0.9    & 3.1  $^{+  1.1   }_{- 1.0  }$  & 4.4 \\ 
AzTEC J221730.02+000439.6 & SSA22-AzTEC61 & 22 17 30.02    &  +00 04 39.6    & 3.3   $\pm$ 0.7    & 2.6  $^{+  0.8   }_{- 0.8  }$  & 4.4 \\ 
AzTEC J221728.42+000730.7 & SSA22-AzTEC62 & 22 17 28.42    &  +00 07 30.7    & 3.2   $\pm$ 0.7    & 2.6  $^{+  0.8   }_{- 0.9  }$  & 4.4 \\ 
AzTEC J221640.23+001026.1 & SSA22-AzTEC63 & 22 16 40.23    &  +00 10 26.1    & 4.0   $\pm$ 0.9    & 3.0  $^{+  1.1   }_{- 1.0  }$  & 4.4 \\ 
AzTEC J221632.06+001742.5 & SSA22-AzTEC64 & 22 16 32.06    &  +00 17 42.5    & 4.4   $\pm$ 1.0    & 3.3  $^{+  1.2   }_{- 1.2  }$  & 4.3 \\ 
AzTEC J221800.83+002930.1 & SSA22-AzTEC65 & 22 18 00.83    &  +00 29 30.1    & 4.6   $\pm$ 1.1    & 3.4  $^{+  1.2   }_{- 1.2  }$  & 4.3 \\ 
AzTEC J221827.80+001151.7 & SSA22-AzTEC66 & 22 18 27.80    &  +00 11 51.7    & 3.4   $\pm$ 0.8    & 2.7  $^{+  0.9   }_{- 0.9  }$  & 4.3 \\ 
AzTEC J221707.38+001317.5 & SSA22-AzTEC67 & 22 17 07.38    &  +00 13 17.5    & 3.1   $\pm$ 0.7    & 2.5  $^{+  0.8   }_{- 0.8  }$  & 4.3 \\ 
AzTEC J221812.36+002320.2 & SSA22-AzTEC68 & 22 18 12.36    &  +00 23 20.2    & 3.6   $\pm$ 0.8    & 2.8  $^{+  0.9   }_{- 1.0  }$  & 4.3 \\ 
AzTEC J221747.05+002233.2 & SSA22-AzTEC69 & 22 17 47.05    &  +00 22 33.2    & 3.1   $\pm$ 0.7    & 2.5  $^{+  0.8   }_{- 0.8  }$  & 4.3 \\ 
AzTEC J221640.34+001417.9 & SSA22-AzTEC70 & 22 16 40.34    &  +00 14 17.9    & 3.5   $\pm$ 0.8    & 2.7  $^{+  0.9   }_{- 0.9  }$  & 4.3 \\ 
AzTEC J221723.62+000929.1 & SSA22-AzTEC71 & 22 17 23.62    &  +00 09 29.1    & 3.1   $\pm$ 0.7    & 2.5  $^{+  0.8   }_{- 0.8  }$  & 4.2 \\ 
AzTEC J221727.84+002850.6 & SSA22-AzTEC72 & 22 17 27.84    &  +00 28 50.6    & 3.6   $\pm$ 0.9    & 2.8  $^{+  0.9   }_{- 1.0  }$  & 4.2 \\ 
AzTEC J221842.03+001659.3 & SSA22-AzTEC73 & 22 18 42.03    &  +00 16 59.3    & 4.2   $\pm$ 1.0    & 3.1  $^{+  1.1   }_{- 1.1  }$  & 4.2 \\ 
AzTEC J221707.16+001725.2 & SSA22-AzTEC74 & 22 17 07.16    &  +00 17 25.2    & 3.1   $\pm$ 0.7    & 2.4  $^{+  0.9   }_{- 0.8  }$  & 4.2 \\ 
AzTEC J221724.22+000723.5 & SSA22-AzTEC75 & 22 17 24.22    &  +00 07 23.5    & 3.1   $\pm$ 0.7    & 2.4  $^{+  0.8   }_{- 0.8  }$  & 4.2 \\ 
AzTEC J221847.75+001417.5 & SSA22-AzTEC76 & 22 18 47.75    &  +00 14 17.5    & 4.7   $\pm$ 1.1    & 3.3  $^{+  1.3   }_{- 1.4  }$  & 4.1 \\ 
AzTEC J221734.86+001539.2 & SSA22-AzTEC77 & 22 17 34.86    &  +00 15 39.2    & 3.2   $\pm$ 0.8    & 2.4  $^{+  0.9   }_{- 0.8  }$  & 4.1 \\ 
AzTEC J221836.10+001006.5 & SSA22-AzTEC78 & 22 18 36.10    &  +00 10 06.5    & 3.9   $\pm$ 1.0    & 2.9  $^{+  1.1   }_{- 1.1  }$  & 4.1 \\ 
AzTEC J221709.16$-$000057.0 & SSA22-AzTEC79 & 22 17 09.16    &  $-$00 00 57.0    & 5.1   $\pm$ 1.3    & 3.4  $^{+  1.5   }_{- 1.5  }$  & 4.1 \\ 
AzTEC J221713.64+001927.5 & SSA22-AzTEC80 & 22 17 13.64    &  +00 19 27.5    & 3.0   $\pm$ 0.7    & 2.3  $^{+  0.9   }_{- 0.8  }$  & 4.1 \\ 
AzTEC J221702.19+001558.6 & SSA22-AzTEC81 & 22 17 02.19    &  +00 15 58.6    & 3.0   $\pm$ 0.7    & 2.3  $^{+  0.9   }_{- 0.8  }$  & 4.1 \\ 
AzTEC J221800.65+002524.7 & SSA22-AzTEC82 & 22 18 00.65    &  +00 25 24.7    & 3.3   $\pm$ 0.8    & 2.5  $^{+  0.9   }_{- 0.9  }$  & 4.1 \\ 
AzTEC J221657.04+002411.5 & SSA22-AzTEC83 & 22 16 57.04    &  +00 24 11.5    & 3.8   $\pm$ 0.9    & 2.7  $^{+  1.1   }_{- 1.0  }$  & 4.0 \\ 
AzTEC J221819.43+002120.9 & SSA22-AzTEC84 & 22 18 19.43    &  +00 21 20.9    & 3.3   $\pm$ 0.8    & 2.5  $^{+  1.0   }_{- 0.9  }$  & 4.0 \\ 
AzTEC J221816.78+002456.6 & SSA22-AzTEC85 & 22 18 16.78    &  +00 24 56.6    & 3.9   $\pm$ 1.0    & 2.8  $^{+  1.2   }_{- 1.1  }$  & 4.0 \\ 
AzTEC J221711.31+000329.2 & SSA22-AzTEC86 & 22 17 11.31    &  +00 03 29.2    & 3.3   $\pm$ 0.8    & 2.5  $^{+  1.0   }_{- 1.0  }$  & 4.0 \\ 
AzTEC J221725.60+002032.9 & SSA22-AzTEC87 & 22 17 25.60    &  +00 20 32.9    & 2.9   $\pm$ 0.7    & 2.3  $^{+  0.8   }_{- 0.9  }$  & 4.0 \\ 
AzTEC J221804.39+002154.0 & SSA22-AzTEC88 & 22 18 04.39    &  +00 21 54.0    & 3.0   $\pm$ 0.8    & 2.3  $^{+  0.9   }_{- 0.9  }$  & 4.0 \\ 
AzTEC J221648.67+001100.2 & SSA22-AzTEC89 & 22 16 48.67    &  +00 11 00.2    & 3.2   $\pm$ 0.8    & 2.4  $^{+  0.9   }_{- 0.9  }$  & 3.9 \\ 
AzTEC J221709.98+001100.5 & SSA22-AzTEC90 & 22 17 09.98    &  +00 11 00.5    & 2.9   $\pm$ 0.7    & 2.2  $^{+  0.9   }_{- 0.8  }$  & 3.9 \\ 
AzTEC J221736.92+000342.9 & SSA22-AzTEC91 & 22 17 36.92    &  +00 03 42.9    & 2.9   $\pm$ 0.8    & 2.2  $^{+  0.9   }_{- 0.8  }$  & 3.9 \\ 
AzTEC J221720.86+000141.0 & SSA22-AzTEC92 & 22 17 20.86    &  +00 01 41.0    & 3.2   $\pm$ 0.8    & 2.4  $^{+  1.0   }_{- 1.0  }$  & 3.9 \\ 
AzTEC J221658.17+000917.7 & SSA22-AzTEC93 & 22 16 58.17    &  +00 09 17.7    & 3.0   $\pm$ 0.8    & 2.3  $^{+  0.9   }_{- 0.9  }$  & 3.9 \\ 
AzTEC J221803.24+000300.6 & SSA22-AzTEC94 & 22 18 03.24    &  +00 03 00.6    & 3.4   $\pm$ 0.9    & 2.5  $^{+  1.0   }_{- 1.0  }$  & 3.9 \\ 
AzTEC J221736.99+000817.5 & SSA22-AzTEC95 & 22 17 36.99    &  +00 08 17.5    & 2.8   $\pm$ 0.7    & 2.2  $^{+  0.8   }_{- 0.9  }$  & 3.9 \\ 
AzTEC J221754.02+001927.4 & SSA22-AzTEC96 & 22 17 54.02    &  +00 19 27.4    & 2.8   $\pm$ 0.7    & 2.2  $^{+  0.8   }_{- 0.9  }$  & 3.9 \\ 
AzTEC J221831.24+001402.7 & SSA22-AzTEC97 & 22 18 31.24    &  +00 14 02.7    & 3.1   $\pm$ 0.8    & 2.3  $^{+  0.9   }_{- 0.9  }$  & 3.9 \\ 
AzTEC J221810.78+001650.5 & SSA22-AzTEC98 & 22 18 10.78    &  +00 16 50.5    & 2.9   $\pm$ 0.7    & 2.2  $^{+  0.8   }_{- 0.9  }$  & 3.9 \\ 
AzTEC J221725.16+001756.8 & SSA22-AzTEC99 & 22 17 25.16    &  +00 17 56.8    & 2.9   $\pm$ 0.7    & 2.2  $^{+  0.8   }_{- 0.9  }$  & 3.9 \\ 
AzTEC J221715.40+001747.9 & SSA22-AzTEC100 & 22 17 15.40    &  +00 17 47.9    & 2.8   $\pm$ 0.7    & 2.2  $^{+  0.8   }_{- 0.9  }$  & 3.8 \\ 
AzTEC J221756.53+002808.6 & SSA22-AzTEC101 & 22 17 56.53    &  +00 28 08.6    & 3.4   $\pm$ 0.9    & 2.5  $^{+  1.0   }_{- 1.1  }$  & 3.8 \\ 
AzTEC J221744.34+000633.1 & SSA22-AzTEC102 & 22 17 44.34    &  +00 06 33.1    & 2.8   $\pm$ 0.7    & 2.1  $^{+  0.9   }_{- 0.8  }$  & 3.8 \\ 
AzTEC J221802.15+000744.0 & SSA22-AzTEC103 & 22 18 02.15    &  +00 07 44.0    & 2.9   $\pm$ 0.8    & 2.2  $^{+  0.9   }_{- 0.9  }$  & 3.8 \\ 
AzTEC J221628.56+001512.6 & SSA22-AzTEC104 & 22 16 28.56    &  +00 15 12.6    & 4.0   $\pm$ 1.0    & 2.7  $^{+  1.3   }_{- 1.3  }$  & 3.8 \\ 
AzTEC J221826.81+002142.5 & SSA22-AzTEC105 & 22 18 26.81    &  +00 21 42.5    & 3.6   $\pm$ 1.0    & 2.5  $^{+  1.1   }_{- 1.1  }$  & 3.8 \\ 
AzTEC J221722.90+001657.8 & SSA22-AzTEC106 & 22 17 22.90    &  +00 16 57.8    & 2.8   $\pm$ 0.7    & 2.1  $^{+  0.8   }_{- 0.9  }$  & 3.8 \\ 
AzTEC J221655.79+000757.7 & SSA22-AzTEC107 & 22 16 55.79    &  +00 07 57.7    & 3.1   $\pm$ 0.8    & 2.3  $^{+  1.0   }_{- 1.0  }$  & 3.8 \\ 
AzTEC J221736.21+002226.9 & SSA22-AzTEC108 & 22 17 36.21    &  +00 22 26.9    & 2.8   $\pm$ 0.7    & 2.1  $^{+  0.8   }_{- 0.9  }$  & 3.8 \\ 
AzTEC J221826.80+002023.9 & SSA22-AzTEC109 & 22 18 26.80    &  +00 20 23.9    & 3.3   $\pm$ 0.9    & 2.3  $^{+  1.1   }_{- 1.0  }$  & 3.7 \\ 
AzTEC J221751.00+001506.4 & SSA22-AzTEC110 & 22 17 51.00    &  +00 15 06.4    & 2.7   $\pm$ 0.7    & 2.1  $^{+  0.8   }_{- 0.9  }$  & 3.7 \\ 
AzTEC J221816.21+002359.9 & SSA22-AzTEC111 & 22 18 16.21    &  +00 23 59.9    & 3.4   $\pm$ 0.9    & 2.4  $^{+  1.0   }_{- 1.1  }$  & 3.7 \\ 
AzTEC J221657.21+001111.6 & SSA22-AzTEC112 & 22 16 57.21    &  +00 11 11.6    & 2.8   $\pm$ 0.8    & 2.1  $^{+  0.9   }_{- 0.9  }$  & 3.7 \\ 
\hline
\end{tabular}
\end{center}
\end{table*}

\setcounter{table}{1}
\begin{table*}
\begin{center}
\caption{-- continued}
\begin{tabular}{ccccccc}
\hline
\hline
Name & ID & R.A. & Dec. & S$_{\rm observed}$ & S$_{\rm deboosted}$ & S/N \\ 
&  & [$^{\rm h}$ $^{\rm m}$ $^{\rm s}$ ] & [$^{\circ}$ $^{\prime}$ $^{\prime\prime}$ ] & [mJy] & [mJy] & \\
 \hline
AzTEC J221748.40+002315.2 & SSA22-AzTEC113 & 22 17 48.40    &  +00 23 15.2    & 2.7   $\pm$ 0.7    & 2.0  $^{+  0.9   }_{- 0.9  }$  & 3.7 \\ 
AzTEC J221645.74+000600.1 & SSA22-AzTEC114 & 22 16 45.74    &  +00 06 00.1    & 4.2   $\pm$ 1.1    & 0.0  $^{+  1.5   }_{- 1.4  }$  & 3.7 \\ 
AzTEC J221706.21+001515.6 & SSA22-AzTEC115 & 22 17 06.21    &  +00 15 15.6    & 2.7   $\pm$ 0.7    & 2.0  $^{+  0.9   }_{- 0.9  }$  & 3.6 \\ 
AzTEC J221719.43+001552.0 & SSA22-AzTEC116 & 22 17 19.43    &  +00 15 52.0    & 2.7   $\pm$ 0.7    & 2.0  $^{+  0.8   }_{- 0.9  }$  & 3.6 \\ 
AzTEC J221732.01+002548.4 & SSA22-AzTEC117 & 22 17 32.01    &  +00 25 48.4    & 2.8   $\pm$ 0.8    & 2.0  $^{+  0.9   }_{- 0.9  }$  & 3.6 \\ 
AzTEC J221759.46$-$000021.4 & SSA22-AzTEC118 & 22 17 59.46    &  $-$00 00 21.4    & 4.0   $\pm$ 1.1    & 2.6  $^{+  1.3   }_{- 1.4  }$  & 3.6 \\ 
AzTEC J221706.83+000225.8 & SSA22-AzTEC119 & 22 17 06.83    &  +00 02 25.8    & 3.4   $\pm$ 0.9    & 2.3  $^{+  1.2   }_{- 1.1  }$  & 3.6 \\ 
AzTEC J221832.78+002254.9 & SSA22-AzTEC120 & 22 18 32.78    &  +00 22 54.9    & 4.5   $\pm$ 1.2    & 0.0  $^{+  1.6   }_{- 1.6  }$  & 3.6 \\ 
AzTEC J221833.42+002114.9 & SSA22-AzTEC121 & 22 18 33.42    &  +00 21 14.9    & 3.9   $\pm$ 1.1    & 0.0  $^{+  1.3   }_{- 1.4  }$  & 3.6 \\ 
AzTEC J221746.04$-$000223.9 & SSA22-AzTEC122 & 22 17 46.04    &  $-$00 02 23.9    & 3.9   $\pm$ 1.1    & 0.0  $^{+  1.3   }_{- 1.4  }$  & 3.6 \\ 
AzTEC J221659.11+002305.5 & SSA22-AzTEC123 & 22 16 59.11    &  +00 23 05.5    & 3.0   $\pm$ 0.8    & 2.1  $^{+  1.0   }_{- 1.1  }$  & 3.5 \\ 
AzTEC J221834.38+001738.1 & SSA22-AzTEC124 & 22 18 34.38    &  +00 17 38.1    & 3.1   $\pm$ 0.9    & 2.1  $^{+  1.1   }_{- 1.1  }$  & 3.5 \\ 
AzTEC J221755.73+001941.3 & SSA22-AzTEC125 & 22 17 55.73    &  +00 19 41.3    & 2.6   $\pm$ 0.7    & 1.9  $^{+  0.9   }_{- 0.9  }$  & 3.5 \\ 
\hline
\end{tabular}
\end{center}
\end{table*}

\begin{table*}
\begin{center}
\begin{tabular}{cccccc}
\hline
\hline
ID & R.A. & Dec. & S$_{\rm observed}$ & S/N \\
 & [$^{\rm h}$ $^{\rm m}$ $^{\rm s}$ ] & [$^{\circ}$ $^{\prime}$ $^{\prime\prime}$ ] & [mJy] &  \\
 
\hline
SSA22-AzTEC-Sup1 & 22 16 56.16    &  +00 28 43.1    & 10.3   & 6.6 \\ 
SSA22-AzTEC-Sup2 & 22 16 33.20    &  +00 07 23.6    & 8.7   & 6.0 \\ 
SSA22-AzTEC-Sup3 & 22 18 23.56    &  +00 26 31.1    & 7.5   & 5.5 \\ 
SSA22-AzTEC-Sup4 & 22 16 22.10    &  +00 10 21.7    & 7.3   & 4.4 \\ 
SSA22-AzTEC-Sup5 & 22 18 07.01    &  -00 03 29.6    & 9.9   & 4.3 \\ 
SSA22-AzTEC-Sup6 & 22 17 16.35    &  -00 03 48.0    & 6.9   & 4.2 \\ 
SSA22-AzTEC-Sup7 & 22 16 34.92    &  +00 22 30.7    & 5.4   & 4.0 \\ 
SSA22-AzTEC-Sup8 & 22 16 23.96    &  +00 21 39.5    & 7.9   & 4.0 \\ 
SSA22-AzTEC-Sup9 & 22 16 44.60    &  +00 01 32.5    & 8.3   & 4.0 \\ 
SSA22-AzTEC-Sup10 & 22 17 08.81    &  -00 02 53.6    & 6.6   & 3.8 \\ 
SSA22-AzTEC-Sup11 & 22 18 20.10    &  +00 02 03.3    & 5.3   & 3.8 \\ 
SSA22-AzTEC-Sup12 & 22 18 46.36    &  +00 06 03.7    & 7.7   & 3.7 \\ 
SSA22-AzTEC-Sup13 & 22 18 52.95    &  +00 15 19.2    & 5.2   & 3.7 \\ 
SSA22-AzTEC-Sup14 & 22 18 51.39    &  +00 10 27.2    & 5.8   & 3.7 \\ 
SSA22-AzTEC-Sup15 & 22 18 56.75    &  +00 14 10.2    & 6.3   & 3.7 \\ 
SSA22-AzTEC-Sup16 & 22 18 24.77    &  +00 28 05.7    & 6.3   & 3.6 \\ 
SSA22-AzTEC-Sup17 & 22 16 38.97    &  +00 05 33.9    & 5.2   & 3.5 \\ 
\hline
\end{tabular}
\end{center}
\caption{Supplementary catalog of AzTEC SMGs detected in the edge (10-30\% coverage) region.}
\end{table*}

The resulting 1.1-mm map is shown in Figure \ref{aztec_map}.  In this paper we consider the area where the weight (i.e., inverse-square of the noise level) is equal to or greater than 30\% of the maximum weight as the survey area in this paper.  The surface area that AzTEC covered was 0.27~deg$^2$, in which the $1\sigma$ noise level ranges from 0.7 to 1.3 mJy~beam$^{-1}$ (50\% of the region has $< 0.8$~mJy~beam$^{-1}$).  The map area corresponds to approximately $60 \times 60$ Mpc$^2$ in a comoving area at $z = 3.1$, which is large enough to cover the protocluster.

The 1.1-mm sources are extracted from the signal-to-noise map, made from the signal map divided by the noise map, with a detection threshold of $\ge 3.5\sigma$.  Each source position is defined by flux-squared weighting of the pixels of the nominal peak within a radius of 15 arcsec. We detected 125 sources in the 30\% coverage area, which are listed in Table 1.  Seventeen $\ge 3.5\sigma$ sources were detected in the 10--30\% coverage area. These are summarized in a supplementary source catalogue (Table 3).

For a source population whose number counts decrease rapidly with increasing flux density, the measured flux of a low signal-to-noise-ratio (S/N) source can be boosted by random noise.
The flux densities of the detected sources are de-boosted to correct for this flux bias using the 
Bayesian recipe described in \cite{2006MNRAS.372.1621C} and
 \cite{2008MNRAS.385.2225S}.  To compute the posterior probability distribution function (PDF) of the intrinsic flux densities, we simulated a prior distribution function and a likelihood function at each position where the actual source is detected. In order to estimate the prior distribution function, we made $10^4$ maps of photon-noise-free random sky realizations that we would observe with the AzTEC/ASTE kernel according to the best-estimate of the 1.1-mm source counts (\citealt{2012MNRAS.423..575S}).
By creating a flux histogram across the maps we arrived at the prior distribution function. We approximated the likelihood function with a normal distribution with $\sigma$ being the local noise level at each source position.  The PDF was obtained by multiplying the prior and likelihood functions. The de-boosted flux is then given by the flux that gives a local maximum of the PDF closest to the measured flux. The de-boosted flux and error bars (68\% confidence intervals) are also listed in Table 2.

\subsection{Characterization of a Map}

  \begin{figure}
    \centering
    \includegraphics[width=0.99\columnwidth]{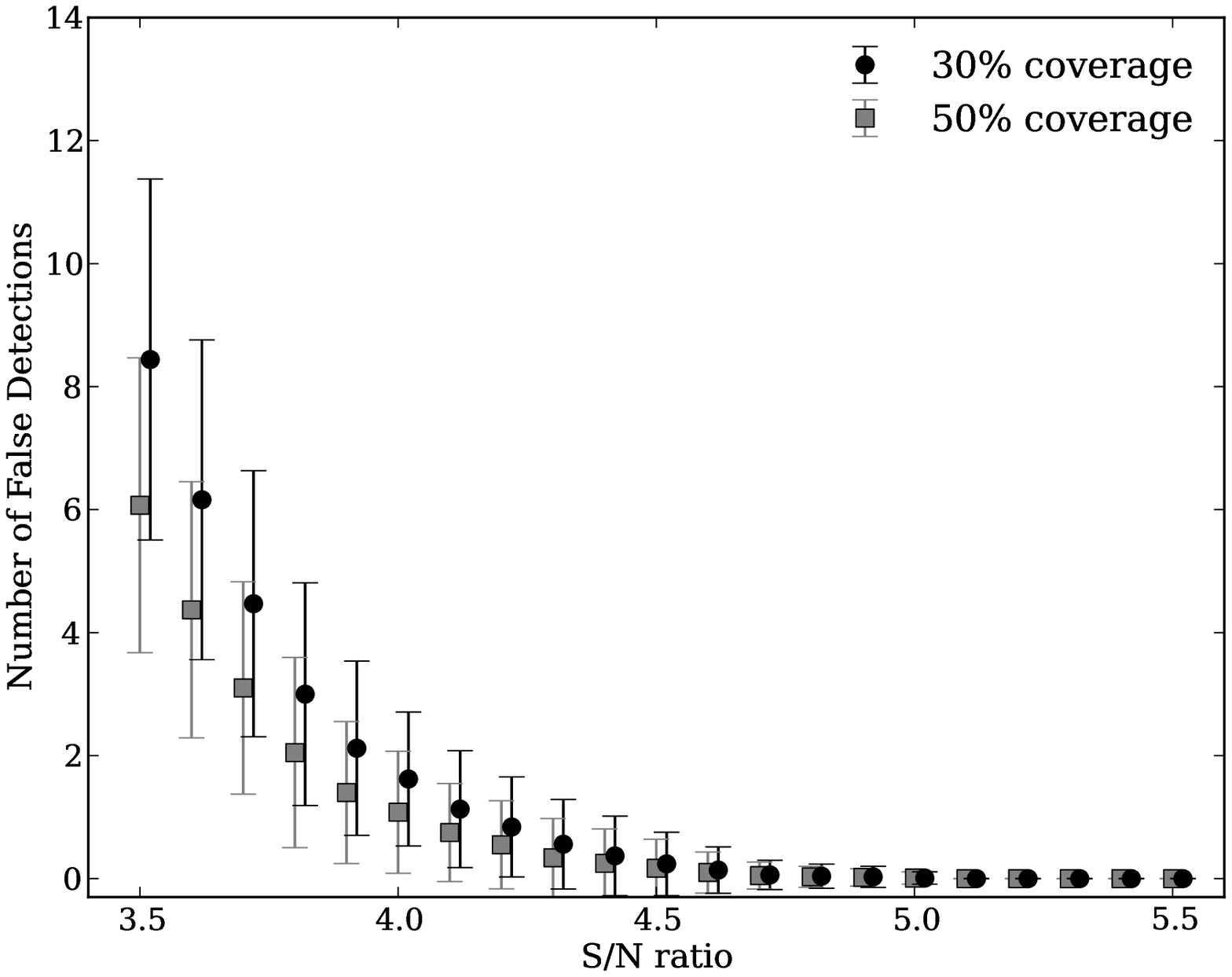}
    \caption{
    Cumulative number count of false detections as a function of the signal to noise ratio.
    The error bars represent 1$\sigma$ Poisson uncertainties.  
    Circles and squares represent the 50 \% and 30 \% coverage field, respectively.
    Data points for the 30 \% coverage field are displaced by $+$0.02 in the S/N ratio for clarity.
    }
    \label{fdr}
  \end{figure}

  \begin{figure}
    \centering
    \includegraphics[width=0.99\columnwidth]{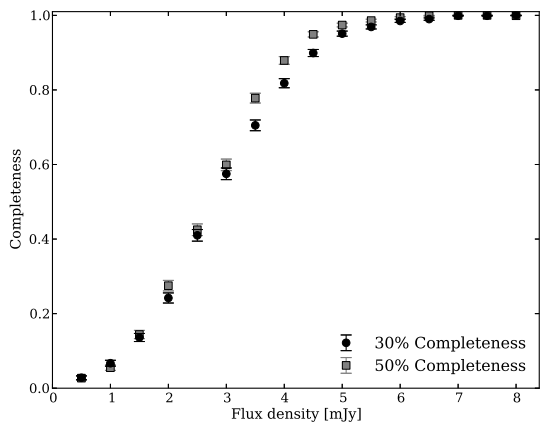}
    \caption{
    Completeness of the AzTEC/ASTE survey in the 50 \% and 30 \% coverage field.
    The error bars represent 1$\sigma$ estimated from the binomial distribution.
    }
    \label{comp}
  \end{figure}

  \begin{figure}
    \centering
    \includegraphics[width=0.99\columnwidth]{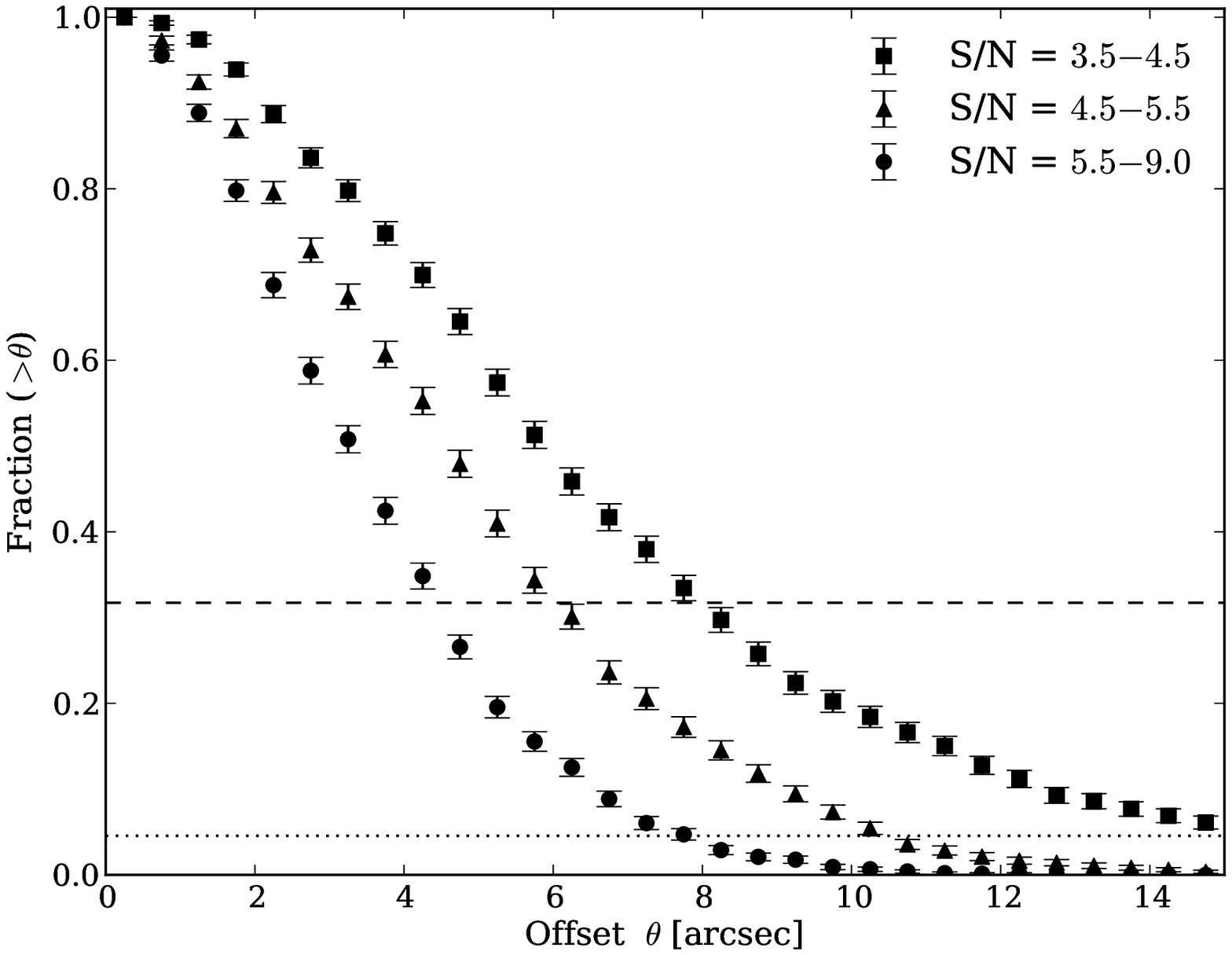}
    \caption{Positional uncertainties estimated from the cumulative probability distribution for a source being detected at a position $\theta$ arcsec away from the true position. The probability was calculated for sources with 3.5$\le$S/N$<$4.5, 4.5$\le$S/N$<$5.5, and 5.5$\le$S/N $<$9.0. The horizontal dashed and dotted lines represent 68.3\% and 99.5\% confidence levels, respectively.}
    \label{pud}
  \end{figure}

\subsubsection{False Detections}

Some fraction of the AzTEC sources were expected to be spurious sources due to positive noise fluctuation, especially when the source had a modest value of S/N. To estimate the number of such non-real sources in the SSA22 map, we extracted sources from a set of jackknifed noise maps. The standard source extraction procedures were performed for 100 simulated maps within the 30\% coverage region.
Figure \ref{fdr} represents the obtained cumulative false detection rate. The computed number of false detections is shown as a function of S/N. At S/N $\approx$ 3.5, approximately 8 out of 125 sources (6\%) were predicted to be spurious.

\subsubsection{Completeness}

The survey completeness is defined as the probability that a real source will result in a measurement above the detection threshold. To evaluate it, we measured the recovery rate of point sources with known flux densities that were embedded into the signal map. The sources were embedded one at the time, using flux densities ranging from 0.5 mJy to 8.0 mJy. The input positions were selected randomly within the 30\% coverage region.
To avoid source blending, the area around from a real source (within 20$^{\prime\prime}$) was not permitted for embedding sources.
If an input source was detected within 20$^{\prime\prime}$ of its embedded position with S/N $\gg$3.5, it was approved to be recovered. We performed 1000 trials and found that the survey completeness was 50\% at a flux density of 2.75 mJy.
Figure \ref{comp} shows the differential completeness as a function of flux density.
The error bars were calculated assuming a binomial distribution.

\subsubsection{Positional Uncertainty}

The detected position of a 1.1 mm source is affected by random and confusion noise in the map, which leads to a large positional error. We estimated such positional uncertainty in a way similar to the one in Section 2.4.2. 
A point source was inserted with known flux density into the SSA22 signal map. We extracted it using a standard algorithm and measured the distribution of input to output source distances as a function of S/N. We repeated these processes for three sample S/N bins (3.5$\le$S/N$<$4.5, 4.5$\le$S/N$<$5.5, and 5.5$\le$S/N $<$9.0).
Figure. \ref{pud} shows the probability that a source detected with a given S/N will be found outside a radial distance $\theta$ from its intrinsic position.

\section{Multi-Wavelength Data and Cataloging}

We utilized optical-to-radio multi-wavelength imaging data to search for counterparts of SMGs discovered by the AzTEC/ASTE survey and estimated the photometric redshift, though most of these observations did not cover the survey area completely, as shown in Figure \ref{aztec_map}.
For counterpart identification we employed the VLA 1.4 GHz, MIPS 24 $\mu$m, and IRAC 3.6 $\mu$m $-$ 8.0 $\mu$m data, all described in more detail below. Then, in addition to the IRAC data, the optical to near-infrared photometry data taken with CFHT/Mega-Cam, Subaru/Suprime-Cam, MOIRCS, and UKIRT were utilized to derive the photometric redshift.

\subsection{VLA 1.4 GHz}

The region centered on the coordinate (RA, Dec)$_{\rm J2000}$ =(22h17m48.0s, +00$^\circ$17$^\prime$13$^{\prime\prime}$) was observed for a total of 48 hours at 1.4 GHz using the National Radio Astronomy Observatory's (NRAO's) Very Large Array (VLA) in its A and B configurations (\citealt{2004ApJ...606...85C}).
Our imaging applied a 50 k$\lambda$ Gaussian taper to reduce the band width smearing.
This produced a well-behaved synthesized beam of 3.0$^{\prime\prime} \times2.9^{\prime\prime}$ at a position angle of $-$80 deg.
Generally, a root-mean-square (r.m.s.) noise level of $\sim8.5$ $\mu$Jy/beam was achieved near the phase center and $\sim20$ $\mu$Jy at 15$^\prime$ from the phase center.
Unfortunately, the radio map was partly contaminated by remarkable side lobes, which were generated by nearby radio-loud sources. Therefore, to avoid misidentification when cataloguing the radio sources we eliminated areas with a local noise level of $\ge$ 20 $\mu$Jy/beam in cataloging radio sources.
The local noise was estimated using the {\it AIPS} task, {\small \sc RMSD}, in a 100 $\times$100 pixel region centered on each pixel.
To generate a source catalog and to measure source properties, we used the task of {\small \sc SAD}.
Finally 40 radio sources were detected with S/N of $\ge 4$ for 66 SMGs.

\subsection{Spitzer/MIPS, IRAC}

In the SSA22 field, several archival {\it Spitzer} near to mid-infrared imaging observations are available and 
IRAC ch1 to ch4 (3.6, 4.5, 5.8, and 8.0 $\mu$m) and MIPS ch1 (24 $\mu$m) images were obtained through the {\it Spitzer} Science Center (SSC) web site.
As shown in Figure. \ref{aztec_map}, the area observed with IRAC and MIPS corresponds to about a half of the area of AzTEC/ASTE.
The FWHM of the PSF was 1.7$^{\prime\prime}$, 1.7$^{\prime\prime}$, 1.9$^{\prime\prime}$, 2.0$^{\prime\prime}$, and 6$^{\prime\prime}$, respectively.
The source catalogs were generated in two ways.
In the case of the MIPS 24 $\mu$m data, we extracted sources with S/N $\ge$ 5 and measured the flux densities using the APEX module within the MOPEX software package ver. 18.4.9 (\citealt{2005PASP..117.1113M}) through point-response-function (PRF) fitting.
For the IRAC data, we utilized {\sc \small SExtractor} ver. 2.8.6 (\citealt{1996A&AS..117..393B}).
We extracted sources that had at least 5 contiguous 0.6$^{\prime\prime}\times0.6^{\prime\prime}$ pixels with fluxes at least 1.5 times the background noise.
Then flux densities were measured in 4.8$^{\prime\prime}$ diameter apertures using the {\sc \small IRAF} (ver. 2.14) task, {\sc \small APPHOT}.
Sources at or above the 3 $\sigma$ corresponding to their local noise were listed up.
The representative depth for each band is summarized in Table 4.
The aperture corrections suggested in the IRAC Instrument Handbook were considered for all IRAC catalogs and the flux errors are expected to be within 10 \% (\citealt{2009ApJ...699.1610H}).

\subsection{Optical to Near-infrared}
Optical imaging observations with Subaru/Suprime-Cam (\citealt{2002PASJ...54..833M}) using five broad band filters ({\it B, V, R, i$^\prime$, and z$^\prime$}) and one narrow band filter (NB497) were obtained by \cite{2004AJ....128.2073H}. 
We also utilized archival CFHT/MegaCam {\it U} band data (P.I. L.Cowie) for the same area.
Just 21 SMGs were also observed with Subaru/MOIRCS (\citealt{2008PASJ...60.1347S}) at {\it J, H}, and {\it Ks} bands (\citealt{2008PASJ...60..683U}, \citealt{2012ApJ...750..116U}).
We made a set of photometry catalogs at these wavelengths utilizing {\sc \small SExtractor}.
We detected sources that had 5 connected pixels above the 2 $\sigma$ noise level.
The flux densities were measured with a 2$^{\prime\prime}$ and 1.1$^{\prime\prime}$ diameter aperture for {\it U, B, V, R, i$^\prime$, z$^\prime$} bands and {\it J, H, Ks} bands respectively.
We then cut the catalogs to $\ge$ 5 $\sigma$, as derived by \cite{2004AJ....128.2073H} and \cite{2008PASJ...60..683U} as listed in Table 4.
In addition, 
the archival $J$ and $K$ band Deep Extragalactic Survey (DXS)  catalogs by UKIRT/Wide Field Camera (WFCAM) (\citealt{2007MNRAS.379.1599L}) are available for all SMGs. 
These data are shallower than MOIRCS data, but valuable for us since they can supplement the shortage of MOIRCS observations at near-infrared wavelengths.

\begin{table}
\begin{center}
\footnotesize
\begin{tabular}{cccccccc} 
\hline
Band &$\lambda _{eff}$ & 5$\sigma$ Depth  & Ref\\
 & ($\mu$m) & (mag)  & \\
\hline
CFHT/MegaCam\\
$Un$ & 0.28 & 26.1 & 1\\
Subaru/Suprime-Cam\\
 $B$ & 0.44 & 26.5  & 2\\
 NB497 & 0.50 & 26.2 & 2\\
 $V$ & 0.54 & 26.6 & 2\\
 $R$ & 0.65 & 26.7 & 2\\
 $i'$ & 0.77 & 26.4  & 2\\
 $z'$ & 0.92 & 25.7 & 2\\
 Subaru/MOIRCS \\
 $J$ & 1.25 & 24.5 &3\\
 $H$ & 1.64 & 24.3 & 3 \\
 $Ks$ & 2.15 & 23.9 &3 \\
UKIRT/WFCAM \\
 $J$ &  1.26 & 23.3 & 4\\
 $K$ &  2.24 & 22.7 & 4\\
 $Spitzer$/IRAC\\
ch1 & 3.56 & 24.9$^a$  & 5 \\
ch2 & 4.51 & 24.1$^a$  & 5\\
 ch3 & 5.76 & 22.0$^a$  & 5\\
ch4 & 7.96 & 21.6$^a$  & 5 \\
 $Spitzer$/MIPS\\
ch1 & 23.68 & 18.3$^a$ & 5\\
\hline
\end{tabular}
\end{center}
\label{obs}
\caption{Summary of optical to near/mid infrared photometry employed in this paper.
The references are as follows.;
1. Archival data, 2. Hayashino et al. (2004), 3. Uchimoto et al. (2012), 4. Lawrence et al. (2007), 5. Hainline et al. (2009)
$^a$ {\small Median formal point-source sensitivity in the combined mosaic of 3$\sigma$ (IRAC) and 5$\sigma$ (MIPS).}}
\end{table}

\section{Counterpart Identification}

The majority of SMGs that were discovered by AzTEC/ASTE in the SSA22 field had no interferometric observations at submm/mm wavelengths.
Only SSA22-AzTEC1 was observed with SMA (\citealt{2010ApJ...724.1270T}). In addition, 
SSA22-AzTEC12 and SSA22-AzTEC77 were confirmed with CO line emission (\citealt{2005ApJ...622..772C}).
Therefore, identifying counterparts in images taken at other wavelengths is an essential process in order to determine their accurate position and to investigate their nature. In this work we adopted the methods outlined by \cite{1986MNRAS.218...31D} and \cite{2007MNRAS.380..199I} as well as some subsequent works (e.g., \citealt{2011MNRAS.415.1479W}, \citealt{2011MNRAS.413.2314B},\citealt{2012MNRAS.420..957Y} , and \citealt{2012MNRAS.426.1845M} ).
We utilized  VLA 1.4 GHz, MIPS 24 $\mu$m, and IRAC 3.6, 4.5, 5.8, and 8.0 $\mu$m imaging data.
These data have better angular resolution than the AzTEC/ASTE survey. At the same time they are expected to detect SMGs with high probability. VLA 1.4 GHz reaches $\theta_{\rm FWHM}\sim 3^{\prime\prime}$.
It is well known that active star forming galaxies radiate remarkable radio emission that reflects the degree of star formation activity (\citealt{1992ARA&A..30..575C}).
MIPS 24 $\mu$m is also suitable to detect SMGs and 
while its angular resolution of $\theta_{\rm FWHM}\sim 6^{\prime\prime}$ is sometimes not enough to identify counterparts alone, it is significantly better than AzTEC/ASTE.
Furthermore 
\cite{2008MNRAS.389..333Y} suggested that SMGs always have a characteristic IRAC color in the ([5.8$\mu$m]/[3.6$\mu$m] vs [8.0$\mu$m/4.5$\mu$m]) diagram and this idea is consistent with a small subset of SMGs observed by SMA.
The IRAC observations achieve $\theta_{\rm FWHM}\sim 2^{\prime\prime}-4^{\prime\prime}$ corresponding to channels.

Sources which lie within a 2$\sigma$ positional uncertainty, i.e., within a radius $R_S$ from an AzTEC/ASTE centroid position were extracted for each data set.
$R_S$s were calculated as functions of S/Ns through Monte Carlo simulations (more details about this process are described in the supplementary information of \citealt{2009Natur.459...61T} and \citealt{2011MNRAS.411..102H}).
Then we calculated the corrected Poisson probability (``$p$-value'', hereafter $p$), the probability of chance association, for all selected candidates. When calculating
$p$, unlike most previous studies, we considered the number density of not brighter than the candidate but all sources in each catalog as well as \cite{2012MNRAS.420..957Y} to avoid underestimating the $p$ of nearby bright radio and/or mid-IR sources.
%(The formulas utilized in calculating $p$ are described in Appendix. B.)
In this paper, we considered sources with $p\leq0.05$ as robust counterparts.
Likewise, we have listed also tentative sources ($0.05 < p \leq 0.20$).
%Optimistic criteria might attribute to increase the fraction of contaminants, whilst a number of real counterparts are expected to be still contained.
In calculating the source surface density, we faced a difficulty in the case of radio data. In order to estimate the $p$ values, we must count what was detected as a source in the actual SSA22 map (i.e., not the real source density). However, terrible side lobes, which are caused by some bright radio objects located near this field, contaminated the 1.4 GHz radio image widely and we could only utilize some patchy regions that had low local noise levels, as mentioned in section 3.1. Thus, it was impossible to estimate source counts for the whole region or for an significant wide submap. Therefore, we sampled small regions randomly across the map to estimate the radio source density. We generated 5000 circle regions having radii of 15$^{\prime\prime}$ and counted radio sources within these regions. Contaminated areas that had local noise levels $> 20~\mu$Jy/beam were excluded.

 \begin{figure}
    \centering
    \includegraphics[width=0.99\columnwidth]{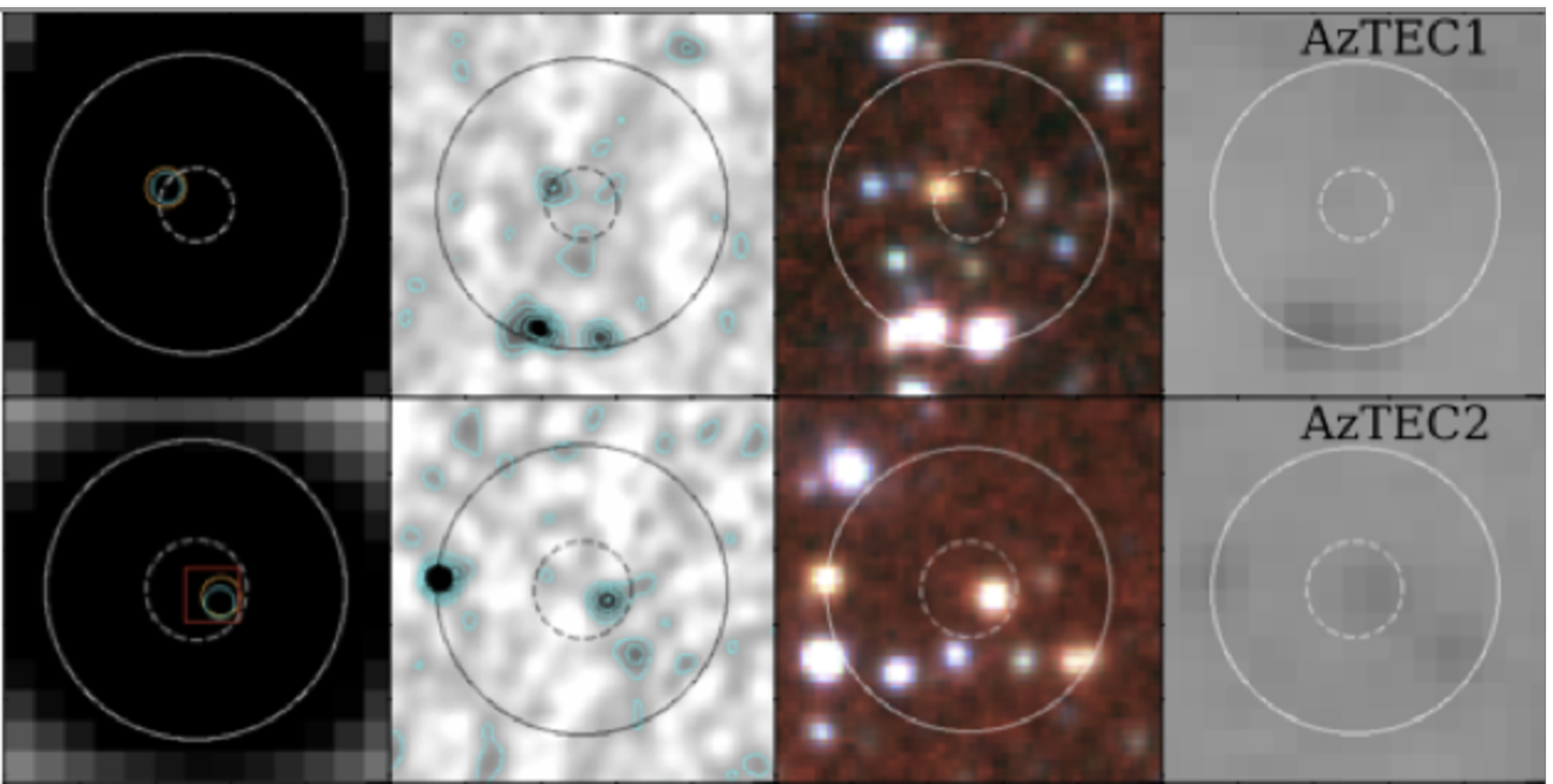}
\caption{Finding charts for the SMGs discovered by AzTEC/ASTE in the SSA22 field.
The full chart is available online.
    In addition to the AzTEC 1.1 mm images, images from VLA 1.4 GHz, IRAC false color, and 24 $\mu$m are shown from left to right.
    The blank or partly lacked  images indicate that these SMGs were not observed at the corresponding wavelengths. 
    The solid and dashed circles show representative FWHM of AzTEC/ASTE (diameter is $30^{\prime \prime}$) and 2$\sigma$ positional error circle. 
    The IRAC color images are produced using the 3.6 $\mu$m (blue), 4.5 $\mu$m (green), and 5.8 $\mu$m (red) band images.
    Each image is 40$^{\prime \prime }\times 40 ^{\prime \prime}$ in size.
    The identified counterparts are shown in the left figure.
    The cyan circles mark the 1.4 GHz radio sources, red squares are the MIPS $24 \mu$m sources. while IRAC sources are represented by orange circles.}
    \label{stamp}
\end{figure}  

%Therefore we estimated the number density of this field through the radio source counts of $Chandra$ deep field south (CDF-S)%(\citealt{2008ApJS..179...71K})
%since their 1.4GHz image has an rms noise of 8.5 $\mu$Jy, which is consistent with our image.
%They obtain the relationship between flux density and source counts as follows:
%$N(>S)=(0.091\pm0.0014)(S/200)^{-1.10\pm0.13}$, 
%where $N$ represent number density of sources with flux density of $>S$ $\mu$Jy
%, measured in units of arcmin$^{-2}$.

\begin{table*}
\caption{Information about identified counterparts. The machine-readable version is available online. 
Robust counterparts are emphasized in boldface.
(SSA22-Az77 was confirmed by CO detection. (Chapman et al. 2005)) 
$Rs$ represent the radius of the 2$\sigma$ positional error circle estimated by Monte Carlo simulations. 
Dist. means the positional offset of counterparts from the AzTEC position.
The {\it Spitzer } coordinate shows the coordinate of IRAC 3.6 $\mu$m as priority.
If counterparts were selected by only MIPS, the MIPS 24 $\mu$m coordinates are shown.
The next three columns after the positional offset show $p$ values of objects selected by VLA 1.4 GHz, $Spitzer$ MIPS 24 $\mu$m, and IRAC color selected objects.
}
\begin{center}
\scriptsize
\begin{tabular}{ccccccccccc}
\hline
ID & $Rs$ & Radio Coordinate & Dist. &{\it Spitzer} Coordinate &  Dist. & $p_{\rm 1.4 GHz}$ & $p_{\rm 24\mu m}$ & $p_{\rm color}$ \\ 
 & ($^{\prime\prime}$) & (J2000) & ($^{\prime\prime}$) & (J2000) & ($^{\prime\prime}$) & & & \\
\hline
\hline
{\bf SSA22-Az1a} & 3.7 &  J221732.41+001743.9 & 3.3 & J221732.42+001743.8 & 3.4 &   {\bf 0.010} & ... &   {\bf 0.042} \\   
{\bf SSA22-Az2a} & 5.1 &  J221742.26+001701.2 & 2.6 & J221742.25+001701.8 & 2.5 &   {\bf 0.011} & {\bf 0.007} &   {\bf 0.045}\\   
{\bf SSA22-Az5a} & 5.7 &  J221718.65+001802.7 & 4.2 & J221718.67+001803.1 & 4.0 &   {\bf 0.020} & {\bf 0.02} &   0.082 & \\   
SSA22-Az6a & 6.0 &  ... & ... & J221815.41+000946.3 & 3.2 & ... & ... &   0.068 & \\   
{\bf SSA22-Az9a} & 6.7 &  J221659.92+001039.0 & 2.5 & ... & ... &   {\bf 0.013} & ... \\   
{\bf SSA22-Az11a} & 7.2 &  ... & ... & J221657.33+001923.9 & 1.3 & ... & ... &   {\bf 0.023} \\   
{\bf SSA22-Az12a} & 7.3 &  ... & ... & J221734.03+001347.5 & 4.1 & ... &   {\bf 0.028} &   0.106 \\   
{\bf SSA22-Az12b} & 7.3 &  J221733.92+001352.2 & 6.8 & J221733.96+001351.9 & 6.2 &   {\bf 0.037} & ... &   0.146 \\   
SSA22-Az13a & 7.5 &  ... & ... & J221720.25+002019.0 & 3.3 & ... & ... &   0.085 \\   
{\bf SSA22-Az14a} & 7.5 &  ... & ... & J221737.02+001821.5 & 0.9 & ... &   {\bf 0.008} &   {\bf 0.013} \\   
SSA22-Az14b & 7.5 &  ... & ... & J221737.22+001816.6 & 6.0 & ... & ... &   0.149 \\   
{\bf SSA22-Az14c} & 7.5 &  J221737.33+001822.9 & 5.6 & ... & ... &   {\bf 0.034} & ... & ... \\   
{\bf SSA22-Az16a} & 7.8 &  ... & ... & J221651.98+001816.9 & 6.0 & ... &   {\bf 0.042} & ... \\   
SSA22-Az18a & 8.1 &  ... & ... & J221742.10+001246.4 & 7.7 & ... & ... &   0.182 \\   
{\bf SSA22-Az20a} & 8.4 &  J221744.06+000822.8 & 2.1 & J221744.08+000822.7 & 1.8 &   {\bf 0.011} & ... &   {\bf 0.042} \\   
{\bf SSA22-Az20b} & 8.4 &  ... & ... & J221744.30+000821.9 & 1.8 & ... & ... &   {\bf 0.042} \\   
SSA22-Az23a & 8.7 &  ... & ... & J221658.21+002452.7 & 6.3 & ... & ... &   0.180 \\   
SSA22-Az23b & 8.7 &  ... & ... & J221658.08+002457.4 & 5.3 & ... & ... &   0.158  \\   
SSA22-Az23c & 8.7 &  ... & ... & J221658.22+002458.1 & 7.3 & ... & ... &   0.197 \\   
SSA22-Az26a & 9.0 &  ... & ... & J221712.99+002654.1 & 6.0 & ... & ... &   0.182 \\   
{\bf SSA22-Az27a} & 9.0 &  ... & ... & J221825.31+001924.8 & 1.3 & ... &   {\bf 0.007} & ... \\   
SSA22-Az28a & 9.1 &  ... & ... & J221650.06+002248.6 & 3.6 & ... & ... &   0.108 \\   
{\bf SSA22-Az30a} & 9.2 &  J221746.77+000804.2 & 6.2 & ... & ... &   {\bf 0.047} & ... & ...  \\   
SSA22-Az31a & 9.3 &  ... & ... & J221727.39+000317.6 & 8.9 & ... &   0.061 &   0.236 \\   
SSA22-Az32a & 9.4 &  ... & ... & J221804.95+000840.1 & 4.7 & ... & ... &   0.150 \\   
SSA22-Az32b & 9.4 &  ... & ... & J221805.26+000839.4 & 2.4 & ... & ... &   0.065 \\   
{\bf SSA22-Az34a} & 9.5 &  ... & ... & J221741.34+002641.4 & 2.9 & ... &   {\bf 0.025} &   0.086 \\   
SSA22-Az34b & 9.5 &  J221741.78+002639.4 & 7.7 & ... & ... &   0.057 & ... & ...\\   
{\bf SSA22-Az35a} & 9.6 &  ... & ... & J221741.27+001045.9 & 5.4 & ... &   {\bf 0.042} &   0.175 \\   
{\bf SSA22-Az35b} & 9.6 &  J221741.44+001045.1 & 3.0 & J221741.30+001045.6 & 4.8 &   {\bf 0.021} &   {\bf 0.042} & ... \\   
SSA22-Az37a & 9.9 &  J221737.39+001024.8 & 7.1 & J221737.39+001024.9 & 7.1 &   0.057 & 0.064 &   0.224 \\   
SSA22-Az45a & 10.7 &  J221732.29+002931.3 & 6.3 & J221732.33+002930.6 & 5.4 &   0.056 & ... &   0.191 \\   
SSA22-Az50a & 11.1 &  ... & ... & J221720.83+002826.1 & 4.8 & ... & ... &   0.172 \\   
SSA22-Az52a & 11.2 &  J221728.32+002025.9 & 10.8 & J221728.37+002027.0 & 9.8 &   0.084 & ... &   0.307 \\   
SSA22-Az52b & 11.2 &  ... & ... & J221728.99+002033.6 & 3.5 & ... & ... &   0.117 \\   
SSA22-Az55a & 11.5 &  ... & ... & J221806.46+001135.4 & 9.5 & ... &   0.094 &   0.314  \\   
{\bf SSA22-Az56a} & 11.8 &  ... & ... & J221806.23+001021.0 & 1.2 & ... & ... &   {\bf 0.025}  \\   
{\bf SSA22-Az59a} & 12.0 &  ... & ... & J221814.32+001446.3 & 4.0 & ... &   {\bf 0.040} & ... \\   
SSA22-Az59b & 12.0 &  ... & ... & J221815.13+001445.4 & 9.3 & ... &   0.096 & ... \\   
SSA22-Az65a & 12.3 &  ... & ... & J221801.22+002928.4 & 5.5 & ... &   0.059 & ... \\   
SSA22-Az67a & 12.3 &  ... & ... & J221707.63+001307.6 & 10.6 & ... &   0.110 &   0.353 \\   
SSA22-Az67b & 12.3 &  ... & ... & J221707.21+001312.9 & 5.1 & ... &   0.065 &   0.199 \\   
SSA22-Az67c & 12.3 &  ... & ... & J221708.10+001323.2 & 12.0 & ... &   0.111 & ... \\
SSA22-Az68a & 12.3 &  ... & ... & J221812.90+002321.4 & 7.6 & ... &   0.085 & ... \\   
SSA22-Az71a & 12.4 &  ... & ... & J221723.13+000926.2 & 7.8 & ... &   0.087 & ... \\   
{\bf SSA22-Az72a} & 12.5 &  J221728.20+002851.4 & 4.9 & J221728.22+002851.7 & 5.2 &   {\bf 0.047} & 0.067 &   0.203 \\   
SSA22-Az74a & 12.5 &  ... & ... & J221706.61+001727.7 & 8.2 & ... &   0.093 &   0.310 \\   
SSA22-Az75a & 12.5 &  ... & ... & J221724.47+000731.7 & 9.1 & ... &   0.106 &   0.332   \\ 
SSA22-Az75b & 12.5 &  ... & ... & J221724.45+000717.5 & 6.9 & ... &   0.078 & ... \\     
SSA22-Az75c & 12.5 &  ... & ... & J221723.79+000714.0 & 11.4 & ... &   0.113 & ... \\   
{\bf SSA22-Az77a} & 12.8 &  ... & ... & J221735.15+001537.4 & 4.6 & ... & ... &   0.180 \\   
{\bf SSA22-Az77b} & 12.8 &  ... & ... & J221734.99+001535.2 & 4.4 & ... &   {\bf 0.046} & ... \\   
SSA22-Az77c & 12.8 &  ... & ... & J221734.98+001527.4 & 12.0 & ... &   0.119 & ... \\   
SSA22-Az80a & 12.9 &  ... & ... & J221713.07+001918.9 & 11.8 & ... &   0.116 &   0.388 \\   
SSA22-Az80b & 12.9 &  ... & ... & J221713.35+001936.6 & 10.0 & ... &   0.117 &   0.361 \\   
SSA22-Az81a & 12.9 &  ... & ... & J221702.29+001550.4 & 8.4 & ... &   0.085 &   0.323 \\   
SSA22-Az81b & 12.9 &  ... & ... & J221702.35+001611.1 & 12.7 & ... &   0.118 &   0.398 \\   
SSA22-Az83a & 13.0 &  ... & ... & J221657.28+002409.6 & 3.9 & ... & ... &   0.148 \\   
SSA22-Az84a & 13.1 &  J221819.43+002126.6 & 5.7 & ... & ... &   0.058 & ... & ... \\   
SSA22-Az85a & 13.1 &  J221816.06+002452.6 & 10.7 & ... & ... &   0.106 & ... & ...  \\   
SSA22-Az85b & 13.1 &  ... & ... & J221816.67+002452.1 & 4.8 & ... &   0.052 & ...  \\   
SSA22-Az87a & 13.2 &  ... & ... & J221726.02+002026.7 & 8.6 & ... &   0.111 &   0.335 \\  
{\bf SSA22-Az88a} & 13.2 &  J221804.43+002153.2 & 0.9 & J221804.49+002153.3 & 1.5 &   {\bf 0.004} &   {\bf 0.009} & ... \\   
SSA22-Az91a & 13.4 &  ... & ... & J221736.64+000343.4 & 4.2 & ... & ... &   0.165 \\   
SSA22-Az95a & 13.5 &  J221736.74+000820.9 & 5.0 & J221736.72+000820.9 & 5.3 &   0.051 & 0.077 &   0.218 \\   
SSA22-Az96a & 13.5 &  ... & ... & J221754.66+001923.3 & 10.0 & ... &   0.121 &   0.375 \\   
SSA22-Az98a & 13.5 &  ... & ... & J221811.25+001656.8 & 9.2 & ... &   0.109 & ... \\   
SSA22-Az99a & 13.5 &  ... & ... & J221725.38+001803.9 & 7.8 & ... &   0.107 &   0.317\\   
SSA22-Az99b & 13.5 &  ... & ... & J221725.18+001805.8 & 9.0 & ... &   0.107 &   0.351 \\   
SSA22-Az99c & 13.5 &  ... & ... & J221724.63+001807.9 & 13.4 & ... &   0.133 & ... \\   
SSA22-Az101a & 13.6 &  ... & ... & J221756.36+002812.9 & 4.8 & ... &   0.054 & ... \\   
SSA22-Az102a & 13.6 &  ... & ... & J221744.48+000632.4 & 2.2 & ... & ... &   0.068 \\   
SSA22-Az103a & 13.7 &  ... & ... & J221801.84+000744.8 & 4.7 & ... & ... &   0.191 \\   
SSA22-Az106a & 13.7 &  J221723.70+001652.4 & 12.8 & J221723.70+001652.4 & 12.8 &   0.123 & 0.136 &   0.428 \\   
\hline
\hline
\end{tabular}
\end{center}
\label{tb_p}
\end{table*}

\setcounter{table}{4}
\begin{table*}
\caption{-- continued.}
\begin{center}
\scriptsize
\begin{tabular}{ccccccccccc}
\hline
SSA22-Az108a & 13.8 &  ... & ... & J221736.48+002219.8 & 8.1 & ... &   0.099 & ... \\   
SSA22-Az109a & 13.9 &  J221826.18+002028.6 & 10.0 & ... & ... &   0.109 & ... \\   
SSA22-Az110a & 13.9 &  J221750.48+001509.3 & 8.1 & J221750.51+001509.4 & 7.6 &   0.090 & ... &   0.318 \\   
SSA22-Az113a & 14.1 &  ... & ... & J221747.60+002318.3 & 11.5 & ... &   0.132 &   0.419 \\   
{\bf SSA22-Az113b} & 14.1 &  ... & ... & J221748.43+002316.6 & 1.4 & ... &   {\bf 0.009} & ... \\   
SSA22-Az115a & 14.2 &  ... & ... & J221705.39+001514.0 & 12.0 & ... &   0.145 &   0.431 \\   
SSA22-Az116a & 14.2 &  ... & ... & J221720.22+001552.9 & 11.4 & ... &   0.134 & ... \\   
SSA22-Az116b & 14.2 &  ... & ... & J221719.34+001545.1 & 7.0 & ... &   0.088 & ... \\    
SSA22-Az117a & 14.2 &  ... & ... & J221732.15+002540.4 & 8.3 & ... &   0.127 &   0.346 \\   
\hline
\hline
\end{tabular}
\end{center}
\label{tb_p}
\end{table*}

We found at least one robust counterpart for 19 AzTEC SMGs.
We also found only tentative counterparts for additional 40 AzTEC SMGs.
The identified objects are listed in Table 5 and the finding charts for the 125 AzTEC SMGs are shown in Figure \ref{stamp} (and online-material). The robust success-rates of identifications based on VLA, MIPS, and IRAC color analysis were 11/66 (17\%), 11/64 (17\%), and 5/61 (8\%), respectively. An additional 10, 26, and 21 AzTEC sources have tentative VLA, MIPS, and IRAC counterparts. The tentative identification rates were 10/66 (15\%), 26/64 (39\%), and 21/61 (32\%), respectively. These rates are generally consistent with previous works for other fields. 
Among them,
\cite{2012MNRAS.420..957Y} provides us with a proper opportunity to compare the results since they also searched counterparts of 1.1-mm-selected SMGs discovered by AzTEC/ASTE in the GOODS-South field and utilized a similar identification method. Their robust (tentative) MIPS and IRAC color identification rates of 8/48(17\%) (14/48, 29\%) and 5/48(10\%) (18/48, 32\%) agree with our studies. In the case of radio identification, their estimates of 13/48 (27\%) for robust sources and 19/48 (40\%) for tentative sources are slightly higher than our rates.
The disparity could be caused by a difference in noise levels of the 1.4 GHz maps since our radio map is noisier than their map.
Indeed, \cite{2012ApJ...761...89B} reported that all 16 SMGs identified by SMA in a SCUBA selected sample, have radio counterparts utilizing the ultra deep ($1\sigma=2.5~\mu$Jy) 1.4 GHz image obtained by JVLA. 
They found that 10 out of 16 these SMGs have the relatively low radio flux density ($S_{\rm1.4 GHz}<40~\mu$Jy).
Therefore surely we missed such faint radio counterparts for a number of SMGs.

Recently submm/mm interferometric observations targeting a number of SMGs have been carried out, providing us with the accuracy rate of these ``traditional methods''.
\cite{2012A&A...548A...4S} made PdBI follow up campaigns targeting 28 APEX/LABOCA selected SMGs in the COSMOS field and found that 
just about  a half of single-dish SMGs have real radio/IR selected counterparts and $>$10\% SMGs should be comprised of multiple sub-mm bright objects.
ALMA observations of 90 SMGs in the CDF-S fields allow us to check answers using more large samples.
They reveal that a number  of predicted counterpart via radio and/or infrared images are identified by ALMA
 ($\sim$ 65\%), though about 40 \% of single dish SMGs have no such predictable counterpart. (\citealt{2013ApJ...768...91H}, \citealt{2013MNRAS.432....2K}).
While these results indicate that statistical approach based on counterpart identification at wavelengths other than (sub-)millimeter is still meaningful, we should note that  our counterpart catalog cannot be correct perfectly.
Additionally,
it has been also reported that a portion of SMGs detected by single-dish telescopes are composed of multiple separated sources.
 For instance, \cite{2013ApJ...768...91H} reported that 24 out of 69 SMGs detected by LABOCA are multiples based on their ALMA observations.
 \cite{2013ApJ...776..131C} showed that 3 out of 24 SMGs identified by SMA in a SCUBA-2 sample are multiples.
 Since the FWHM of the AzTEC/ASTE beam size (30$^{\prime\prime}$) is about 56\% larger than that of LABOCA (19.4$^{\prime\prime}$) and about twice that of SCUBA-2 (14$^{\prime\prime}$), 
multiplicity can be an important factor.

\begin{table}
\begin{center}
\footnotesize
\begin{tabular}{ccc} 
\hline
\hline
Class & Number& Percentage \\
 &  &(\%) \\
\hline
 Total detected SMGs (S/N$\ge3.5$) & 125 &  \\
 SMGs with any of the counterparts & 59 & \\
\hline
SMGs covered by VLA & 66/125 &  53 \\
(with robust VLA counterparts) & 11/66 & 17 \\
(with tentative VLA counterparts) & 10/66 & 15 \\
SMGs covered by MIPS & 64/125 & 51\\
(with robust MIPS counterparts) & 11/64 & 17\\
(with tentative MIPS counterparts) & 26/64 & 39 \\
SMGs covered by IRAC  & 61/125 & 49\\
(with robust IRAC counterparts) & 5/61 & 8\\
(with tentative IRAC counterparts) & 31/61 & 51\\
(with photo-$z$) & 45/61 & 74\\
(without any counterpart) & 14/61 & 23\\
\hline
\end{tabular}
\end{center}
\label{map_prop}
\caption{
SMG identification statistics.
The number of SMGs having counterparts are listed up.
Combining the three data sets, eventually 59 SMGs have at least one of counterparts. 
SMGs with photo-$z$ are composed of IRAC counterparts and VLA/MIPS counterparts covered by IRAC (i.e., some of them have not good $p$ in the case of IRAC).  
Two SMGs (SSA22-AzTEC71 and SSA22-AzTEC108) have MIPS counterparts, but we cannot find corresponding IRAC sources.
}
\end{table}

\section{Photometric Redshift}

For all AzTEC SMGs in the SSA22 field, 
spectroscopic redshifts, $z_{spec}$, are available for only a small subset (5/125,~4\%).
Here we estimate photometric redshifts of reliable (i.e., both of robust and tentative) SMG counterparts to derive redshift distribution and extract candidate SMGs which lie within the $z=3.1$ protocluster.
(\cite{2005ApJ...622..772C} reported that there were 10 SCUBA SMGs which have spectroscopic redshifts in this field.
However only 3 out of 10 sources were detected in the AzTEC map with S/N of $\ge$ 3.5.
The origin of this inconsistency is still unclear, but incompleteness in the AzTEC map might account for this difference.
Indeed most of these sources significantly detected by SCUBA are marginally detected with the AzTEC/ASTE survey (2.5$<$S/N$<$3.5).
Hereafter we mainly focus on AzTEC SMGs.)

For radio counterparts, their photometric redshifts based on 1.1~mm and radio fluxes were estimated with a Bayesian technique (\citealt{2007MNRAS.379.1571A}).
The results were summarized in Table 7.
Derived redshifts were not restricted very much and therefore we mainly discuss redshifts of counterparts based on the optical to near-infrared photometries. 

\subsection{SED fitting}

  \begin{figure}
    \centering
    \includegraphics[width=0.90\columnwidth]{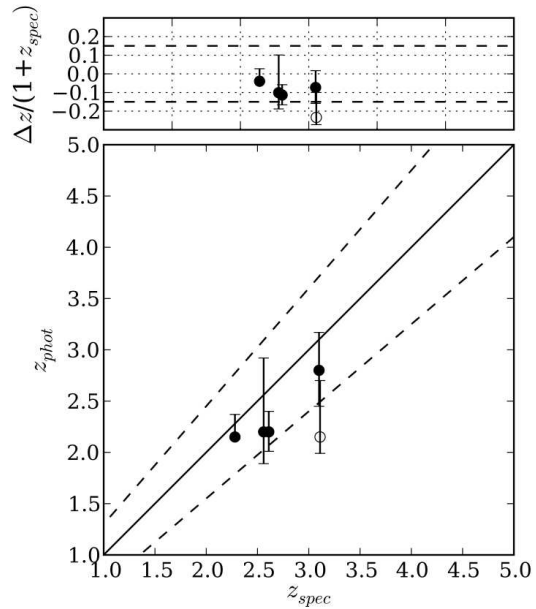}
    \caption{
    Photometric redshifts versus the spectroscopic redshifts from previous works (Matsuda et al. 2005, Chapman et al. 2005, Alaghband-Zadeh et al. 2012.).
The error bars correspond to a $1\sigma$ confidence range.
The dashed lines represent $z_{phot}=z_{spec}\pm0.15(1+z_{spec})$.
Filled circles show sources that were identified as SMGs.
The open circles represent SSA22-Az13a, which has a spectroscopic redshift of 3.1 but is only the most reliable of the tentative counterparts of SSA22-AzTEC13 (see Section 5.2).
The upper panel stands for the errors of ($z_{photo}-z_{spec})/(1+z_{spec})$.
One outlier at $z=3.1$, SSA22-Az13a, is an X-ray source and should be an AGN host (Lehmer et al. (2009), Tamura et al. (2010)). }
    \label{fig:z_comp}
  \end{figure}

We performed SED fitting and obtained photometric redshifts using the {\it HYPERZ} ver. 1.1 code (Bolzonella et al. 2000).
As shown in Figure. \ref{aztec_map},  these imaging area does not cover the AzTEC map completely and a significant part of SMGs does not have complete photometric data. This constraint forced us to consider only the area observed by all IRAC channels when deriving photometric redshifts, since these wavelength data are essential to observe characteristic features such as the 1.6$~\mu$m bump, which is caused by a minimum in the opacity of the negative hydrogen ion in the atmospheres of cool stars (e.g., \citealt{2010ApJ...721.1056S}), and to obtain well-restricted redshift information. It is also difficult to derive photometric redshifts for some radio and/or MIPS counterparts that had no IRAC component. 

Among these 61 SMGs, 14 SMGs have no reliable counterpart concerning all three identification methods.
Two SMGs have MIPS counterparts, but we cannot find corresponding IRAC sources with red color.
Therefore the fitting analysis was performed for the remaining 45 SMGs.
The photometric catalogs of the $U$(CFHT), $B, V, R, i', z' $-band (Subaru/Suprime-Cam), $J, H, Ks$-band (Subaru/MOIRCS), 3.6, 4.5, 5.8, 8.0 $\mu$m(Spitzer/IRAC) were considered, while the $J, K$ (UKIRT) catalogs were utilized only for sources that were not observed by MOIRCS.
We matched the counterpart catalog with the optical to near-infrared catalogs, using a search radius of $r$=1.2 arcsec.
In fitting SEDs, the following parameters were considered.
The range of redshift was set to 0 to 6. 
Dust extinction is considered with a range of $A\rm{v}=0-5$ mag using a bin of 0.5 and we adopted the model of \cite{2000ApJ...533..682C}.
\cite{1993ApJ...405..538B} SED templates of elliptical, Sb, Burst, constant, star-formation (Im) are utilized.

\subsection{Results}

\begin{table*}
\caption{
Redshifts of identified counterparts.
Derived photometric redshifts, $z_p$, with the 68\% confidence intervals based on optical to near-infrared data are shown.
Next column stands for spectroscopic redshifts, $z_s$.
The third column, $z_p$(mm/radio), is photometric redshifts with the 68\% confidence intervals evaluated using 1.1~mm and radio fluxes (We had not calculated redshifts for SSA22-Az106a and SSA22-Az110a, as these were clearly radio-loud AGN and 
difficult to apply the code).
References are as follows: a) $z_p=3.19_{-0.35}^{+0.26}$ in Tamura et al. 2010 , b) Alaghband-Zadeh et al., 2012, c) Chapman et al. 2005, d) Matsuda et al. 2005
}
\begin{center}
\scriptsize
\begin{tabular}{ccccccccc}
\hline
ID & $z_p$ & $z_s$ & $z_p$(mm/radio) & & ID & $z_p$ & $z_s$ & $z_p$(mm/radio) \\ 
\hline
\hline
{\bf SSA22-Az1a} & 2.85 $_{- 0.98 }^{+ 3.15 }$ $^a$ & ... &1.7$_{-0.8}^{+1.8}$ & & {\bf SSA22-Az56a} &  0.95 $_{- 0.07 }^{+ 0.07 }$ & ... & ... \\   
{\bf SSA22-Az2a} & 2.15 $_{- 0.21 }^{+ 0.23 }$ & 2.278$^b$ & 2.1$_{-1.2}^{+1.4}$ & & SSA22-Az67a & 1.70 $_{- 0.30 }^{+ 1.04 }$ & ... & ...\\   
{\bf SSA22-Az5a} & 3.00 $_{- 1.67 }^{+ 1.27 }$ & ... & 1.7$_{-1.0}^{+1.8}$ & & SSA22-Az67b & 1.45 $_{- 0.11 }^{+ 0.30 }$ & ... & ...\\     
SSA22-Az6a & 2.40 $_{- 0.82 }^{+ 3.60 }$ & ... & ... & & {\bf SSA22-Az72a} & 4.10 $_{- 0.26 }^{+ 0.15 }$ & ...& 2.0$_{-1.1}^{+1.5}$\\   
{\bf SSA22-Az9a} & ... & ....& 2.0$_{-0.9}^{+3.5}$ & & SSA22-Az74a & 0.95 $_{- 0.20 }^{+ 0.12 }$ & ... & ... \\   
{\bf SSA22-Az11a} & 2.55 $_{- 0.69 }^{+ 3.45 }$ & ... & ... & & SSA22-Az75a & 1.10 $_{- 0.17 }^{+ 0.26 }$ & ... & ... \\   
{\bf SSA22-Az12a}  & 1.90 $_{- 0.35 }^{+ 0.26 }$ & ... & ...  & & SSA22-Az75c &  0.65 $_{- 0.07 }^{+ 0.16 }$ & ... & ...\\   
{\bf SSA22-Az12b} & 2.20 $_{- 0.32 }^{+ 0.72 }$ &  2.555$^c$& 2.0$_{-1.0}^{+1.6}$ & & {\bf SSA22-Az77a} & 2.80 $_{- 0.34 }^{+ 0.37 }$ & 3.098$^c$ & ...\\   
SSA22-Az13a & 2.15 $_{- 0.16 }^{+ 0.55 }$ & 3.11$^d$ & ... & & SSA22-Az77c & 2.55 $_{- 0.20 }^{+ 0.18 }$ & ... & ...\\   
{\bf SSA22-Az14a} & 2.90 $_{- 0.32 }^{+ 0.21 }$ & ... & ... & & SSA22-Az80a & 0.40 $_{- 0.06 }^{+ 0.05 }$ & ... & ...\\   
SSA22-Az14b & 2.00 $_{- 0.29 }^{+ 0.20 }$ & ... & ... & & SSA22-Az80b &  2.90 $_{- 0.54 }^{+ 0.24 }$ & ... & ...\\   
{\bf SSA22-Az14c} & 3.70 $_{- 0.19 }^{+ 0.72 }$ & ...& 1.8$_{-0.9}^{+2.2}$ & & SSA22-Az81a &2.05 $_{- 0.20 }^{+ 0.32 }$ & ... & ...\\     
SSA22-Az18a & 6.00 $_{- 4.06 }^{+ 0.00 }$ & ... & ... & & SSA22-Az81b & 0.45 $_{- 0.05 }^{+ 0.03 }$ & ... & ...\\   
{\bf SSA22-Az20a} & 2.25 $_{- 0.40 }^{+ 0.20 }$ & ... & 2.0$_{-0.5}^{+3.2}$ & & SSA22-Az83a & 2.55 $_{- 0.10 }^{+ 0.09 }$ & ... & ... \\   
{\bf SSA22-Az20b}  & 3.75 $_{- 0.72 }^{+ 0.21 }$ & ... &... & & SSA22-Az84a & ... & ... & 1.8$_{-1.3}^{+1.3}$\\   
SSA22-Az23a  & 6.00 $_{- 4.18 }^{+ 0.00 }$ & ... & ... & & SSA22-Az85a & ... & ...& 2.1$_{-1.2}^{+1.4}$\\   
SSA22-Az23b & 1.25 $_{- 0.11 }^{+ 0.70 }$ & ... & ... & & SSA22-Az87a & 1.80 $_{- 0.47 }^{+ 0.44 }$ & ... & ...\\   
SSA22-Az23c & 2.35 $_{- 1.41 }^{+ 2.32 }$ & ... & ...  & & {\bf SSA22-Az88a} & ... & ...& 1.9$_{-1.1}^{+1.6}$\\   
SSA22-Az26a & 1.05 $_{- 0.08 }^{+ 0.12 }$ & ... & ...  & & SSA22-Az91a & 2.10 $_{- 0.53 }^{+ 3.42 }$ & ... & ...\\   
SSA22-Az28a & 0.00 $_{- 0.00 }^{+ 0.01 }$ & ... & ... & & SSA22-Az95a & 6.00 $_{- 0.68 }^{+ 0.00 }$ & ... & 2.0$_{-1.5}^{+1.2}$\\   
{\bf SSA22-Az30a} & ... & ... & 1.8$_{-1.3}^{+1.2}$ & & SSA22-Az96a & 2.55 $_{- 0.20 }^{+ 0.08 }$ & ... & ...\\   
SSA22-Az31a & 2.35 $_{- 1.18 }^{+ 2.59 }$ & ... & ... & & SSA22-Az99a & 1.90 $_{- 0.35 }^{+ 0.17 }$ & ... & ... \\   
SSA22-Az32a &  2.25 $_{- 0.91 }^{+ 3.75 }$ & ... & ... & & SSA22-Az99b & 2.25 $_{- 0.30 }^{+ 0.84 }$ & ... & ... \\   
SSA22-Az32b  & 0.85 $_{- 0.07 }^{+ 0.05 }$ & ... & ... & & SSA22-Az102a &  0.65 $_{- 0.07 }^{+ 0.08 }$ & ... & ...\\   
{\bf SSA22-Az34a} & 3.30 $_{- 0.21 }^{+ 0.19 }$ & ...& 1.9$_{-0.9}^{+1.8}$ & & SSA22-Az103a & 1.95 $_{- 0.04 }^{+ 0.04 }$ & ... & ... \\   
{\bf SSA22-Az35a} & 0.35 $_{- 0.06 }^{+ 0.10 }$ & ... & ... & & SSA22-Az106a & 0.45 $_{- 0.05 }^{+ 0.04 }$ & ... & ---\\   
{\bf SSA22-Az35b}  & ... &....& 1.7$_{-0.5}^{+1.5}$ & & SSA22-Az108a & 2.40 $_{- 0.95 }^{+ 0.20 }$ & ... & ...\\   
SSA22-Az37a & 2.20 $_{- 0.25 }^{+ 0.18 }$ & 2.614$^c$& 1.9$_{-1.4}^{+1.2}$ & & SSA22-Az109a & ... & ...& 2.0$_{-1.5}^{+1.5}$\\   
SSA22-Az45a & 1.75 $_{- 0.82 }^{+ 4.25 }$ & ... & 2.1$_{-1.1}^{+1.7}$ & & SSA22-Az110a & 2.75 $_{- 0.80 }^{+ 2.75 }$ & ... & ---\\   
SSA22-Az50a & 1.95 $_{- 0.07 }^{+ 0.37 }$ & ... & ... & & {\bf SSA22-Az113b} & 1.20 $_{- 0.09 }^{+ 0.06 }$ & ... & ...\\   
SSA22-Az52a &...& ... & 1.9$_{-0.9}^{+1.7}$ & & SSA22-Az115a & 2.80 $_{- 0.03 }^{+ 0.07 }$ & ... & ...\\   
SSA22-Az52b & 2.85 $_{- 0.69 }^{+ 2.70 }$ & ... & ... & & SSA22-Az116a &  4.70 $_{- 0.19 }^{+ 0.24 }$ & ... & ... \\   
SSA22-Az55a & 2.20 $_{- 0.05 }^{+ 0.06 }$ & ... & ... & & SSA22-Az117a & 1.30 $_{- 0.26 }^{+ 0.11 }$ & ... & ... \\   
\hline
\hline
\end{tabular}
\end{center}
\label{tb_p}
\end{table*}

We summarize the derived photometric redshifts in Table 7.
%The best-fit SEDs and reduced chi-square values for each case are compiled in Appendix A.
The comparison of spectroscopic and photometric redshifts for SMG counterparts are shown in Figure \ref{fig:z_comp}.
A majority of sources, except for Az13a have comparatively small relative errors.
Only Az13a has a catastrophic error (i.e., $|z_{photo}-z_{spec}|/(1+z_{spec})$ of $>0.15$; e.g.,~\citealt{2009ApJ...690.1236I}).
This source is bright in the X-ray and is supposed to harbor an AGN.
Thus, as a whole, the photometric redshifts are consistent with spectroscopic redshift, though only five (candidate) counterparts have spectroscopic redshifts.

Figure \ref{fig:z_hist} represents the redshift distribution of 48 counterparts of 45 AzTEC SMGs in the SSA22 field. 
All robust counterparts that have photometric or spectroscopic redshifts were considered. 
Three SMGs have two robust counterparts and we utilized both in such cases.
For SMGs that have only tentative counterparts, we selected the most reliable counterpart, i.e., the one with the best $p_{color}$ value.
In calculating the distribution of redshifts, the spectroscopic redshifts were considered when provided by the sources. In the case of SSA22-AzTEC1a, we adopted the photometric redshift using mid and far-infrared data (\citealt{2010ApJ...724.1270T}).
For the remaining sources without spectroscopic redshifts, we considered photometric redshifts. In the same figure two redshift distributions, obtained previously based on optical to near-infrared data, are also shown. One is the SCUBA SMGs that were identified by radio imaging and have spectroscopic redshifts (\citealt{2005ApJ...622..772C}) and the other is composed of AzTEC SMGs in the GOODS-S field with optical photometric redshifts (\citealt{2012MNRAS.420..957Y}).

We derived a median redshift of $z_{med}=$ 2.4 for AzTEC SMGs in the SSA22 field.
It is likely that the distribution of AzTEC SMGs in the SSA22 fields is similar to the ones in the previous two works, though the peak of the distribution is slightly higher than both the SCUBA sources and AzTEC SMGs in the GOODS-S. The shift can be reasonably explained by considering that the longer-wavelength surveys are at an advantage to detect higher redshift SMGs and that the SCUBA SMGs reported in \cite{2005ApJ...622..772C} were composed of radio source only.
Likewise \cite{2012A&A...548A...4S} reported more reliable statistics of interferometrically identified SMGs.
They performed 1.3mm continuum survey with PdBI of 1.1mm-selected SMGs and derived the redshift distribution of these identified counterparts (their redshifts are also generally derived from optical photometry).
The median redshift of 1.1mm-selected SMGs which have photometric redshifts (i.e., not lower limits) is $z_{med}=$ 2.5, which is consistent with our work.
However we should note that these distributions might not represent the whole redshift distribution of AzTEC SMGs. 
It is found that 14 out of 61 SMGs with four-band IRAC coverage (23\%) have no reliable counterparts for all diagnostics, which prevents us from estimating their redshifts completely.
Some of the SMGs are possibly too faint to detect at such wavelengths since they are at 
higher redshift ($z\ge4$).

  \begin{figure}
    \centering
    \includegraphics[width=0.99\columnwidth]{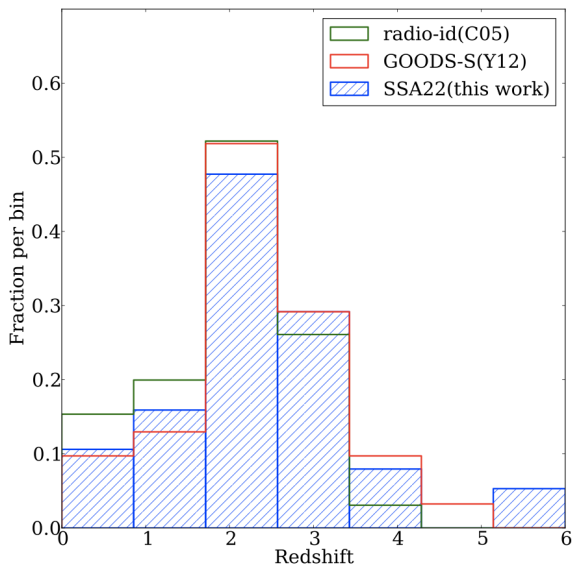}
    \caption{
    Redshift distribution of AzTEC SMGs in the SSA22 field based on photometric/spectroscopic redshifts.
    48 counterpart candidates were considered and $z_{med}=2.4$.
    For comparison, we include 36 AzTEC SMGs with optical photometric redshifts (Yun et al. 2012) and 76 radio-identified SCUBA SMGs with spectroscopic redshifts (Chapman et al. 2005).}
    \label{fig:z_hist}
  \end{figure}

\section{Large Scale Structure}

In the SSA22 field, AzTEC/ASTE covered the area where the large-scale structure traced by $z=3.1$ LAEs was discovered by \cite{2004AJ....128.2073H}.
It is expected that a certain fraction of our SMGs found towards the high-density region of LAEs can reside within the large-scale structure.
Our estimate of the photometric redshifts allows us to extract candidate SMGs at $z=3.1$ and to investigate the relationship between SMGs and the large scale structure.

Combining our estimate of the photometric redshifts with spectroscopic and photometric redshifts} from literature, 
the following 10 AzTEC SMGs are listed up as cluster member candidates.
Firstly, it has been reported that two SMGs should be at $z=3.1$.
Az77a was also detected by the SCUBA survey (SMM J221735.15+001537.2). It had CO detection and its $z_{spec}$=3.098 (\citealt{2005ApJ...622..772C}).
An SMA follow-up observation revealed that Az1a is the real counterpart of SSA22-AzTEC1.
\cite{2010ApJ...724.1270T} estimated its redshift to $z_{phot}=3.19^{+0.26}_{-0.35}$ based on SED fitting in the wavelength range of mid-infrared to radio wavelength range.
Secondly, Az13a is not only an IRAC red object, but also a Lyman alpha emitter, and has a spectroscopic redshift of $z_{spec}=3.11$ (\citealt{2005ApJ...634L.125M}).
Thirdly, we extract 7 sources that could be at $z=3.1$ based on photometric redshifts.
The uncertainty in the photometric redshifts at $z\simeq3$ was estimated as $\Delta z \leq 0.5$ in the case of our data set (\citealt{2012ApJ...750..116U}) and 
therefore we listed up Az5a, Az14a, Az34a, Az52b, Az80b, Az110a and Az115a since they have the most reliable photometric redshift in the range of $z_{phot}=3.1\pm0.5$.
Here we exclude sources that were detected in IRAC channels only (i.e., too faint in the optical range) since we cannot derive well-defined photometric redshifts in such cases.

\subsection{Sky distribution}

  \begin{figure*}
    \centering
    \includegraphics[width=1.5\columnwidth]{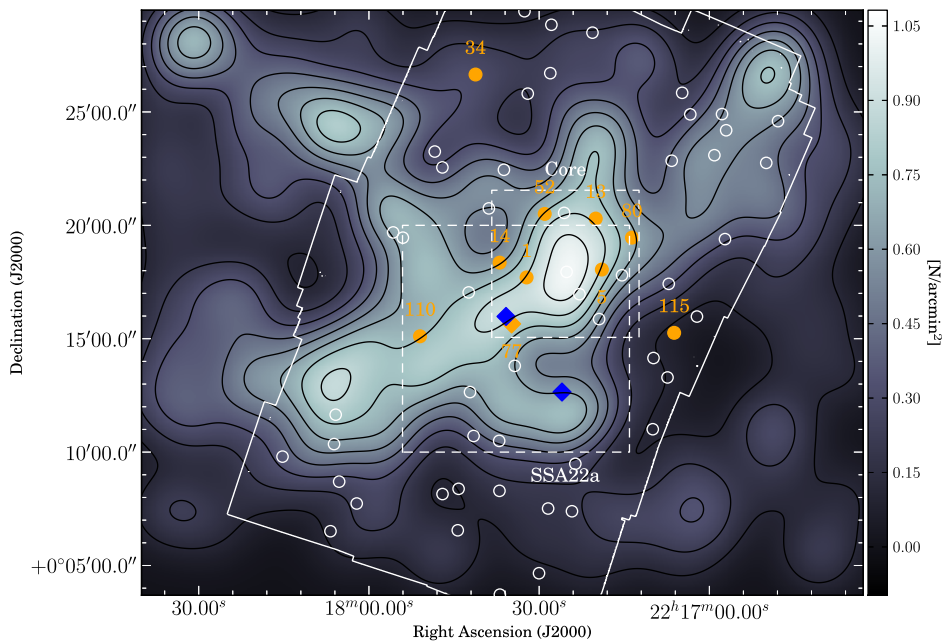}
    \caption{Sky distribution of the $z=3.1$ candidate SMGs and the large scale structure traced by LAEs at $z=3.1$.
    The background map shows the density distribution of $z=3.1$ LAEs (Hayashino et al. 2004). 
    The contours show the smoothed densities of 0.12, 0.23, 0.35, 0.46, 0.57, 0.69, 0.81, and 0.92 galaxies per arcmin$^2$.
    Superposed orange diamonds (circles) show the $z=3.1$ (candidate) AzTEC SMGs. 
   Diamonds are SCUBA SMGs that lie at $z=3.1$.
    The two SCUBA SMGs with blue diamonds were not detected in the AzTEC map (S/N$<$3.5).
    SSA22-AzTEC77 was detected by both of AzTEC and SCUBA and therefore shown as the orange diamond.
    White solid line stands for the area of IRAC ch1 observations.
    White circles denote the remaining AzTEC SMGs located within the IRAC area.
     Seven out of 10 $z\sim3.1$ AzTEC SMGs are concentrated in the core region within the 12 Mpc $\times$ 12 Mpc square (comoving scale).
     For reference we also show the area, ``SSA22a'', named by Steidel et al. (2000).}
    \label{fig:skydis}
  \end{figure*}  

Figure \ref{fig:skydis} shows the two dimensional distribution of the $z=3.1$ candidate SMGs compared with the surface density distribution of LAEs at $z=3.1$ (\citealt{2004AJ....128.2073H}).
The area shown in Figure \ref{fig:skydis} corresponds to a field of view of Suprime-Cam that was called as ``sb1" in \cite{2012AJ....143...79Y}. 
As is evident from this figure, 
the $z=3.1$ candidate SMGs 
are concentrated into the overdense region, while the other SMGs are uniformly distributed across the area observed by IRAC.
Especially 7 SMGs are concentrated to the ``core" region (shown in Figure \ref{fig:skydis}) within the central 12 Mpc $\times$ 12 Mpc (co-moving), which is the most overdense region of LAEs.

It has also been reported that the density peak of the K-band selected (K-selected) galaxies ($K_{AB} < 24$) at $2.6< z_{phot}<3.6$ is consistent with that of LAEs (\citealt{2013ApJ...778..170K}).
This supports that these SMGs are at the center of the protocluster.
They also found both of dusty star-forming galaxies and quiescent galaxies around the ``core" region.
The high source surface density of dusty K-selected galaxies within the region supports the reality of the structure traced by SMGs and 
also implies a high star-forming activity in this overdense region.
%Likewise coexistence of SMGs and massive quiescent galaxies may indicate different stage 

As shown in Figure \ref{fig:skydis}, the ``core" region are roughly corresponds to the area named as SSA22a in \cite{2000ApJ...532..170S}.
\cite{2012AJ....143...79Y} calculated the degree of overdensity of LAEs for sb1 and SSA22a and furthermore estimated the underlying mass fluctuation assuming the standard $\Lambda$CDM model for both regions.
They concluded that 10 and 19 times of the average mass fluctuation was derived, respectively.
Hence SMGs are likely to concentrate in an area where the underlying mass fluctuation is significant high.
Note that this spatial correlation between LAEs and SMGs could not occur accidentally.
When 10 sources are randomly scattered in the IRAC field, the probability that more than 3 sources are located within the core region is $\ll$ 1\%.

Regarding the extents of SMGs, there is an intriguing work, \cite{2012MNRAS.426.1845M}, which reported that there is a clear inconsistency between the redshift distributions of SMGs in the Lockman Hole and UDS fields.
They insisted that the large-scale structures in the $z= 0-5$ universe extending from $\sim 0.3$ to 0.7 deg could explain the difference reasonably.
The extent of the area shown in Figure \ref{fig:skydis} is $\sim 0.5$ deg which correspond to $\sim$ 50 Mpc at $z=3.1$.
Therefore our results generally agree with theirs and indicate that the distribution of SMGs could be patchier than their estimate, at least at $z\sim3$. 

 \begin{figure}
    \centering
    \includegraphics[width=0.99\columnwidth]{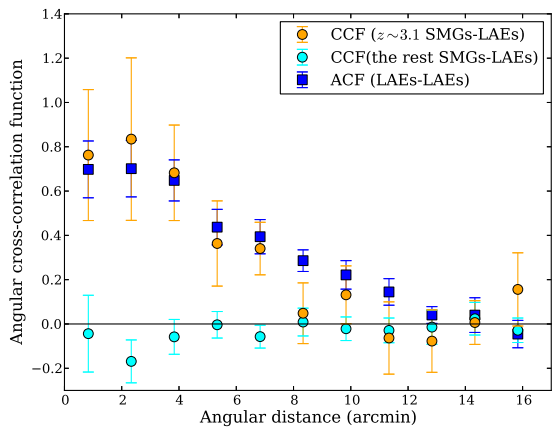}
\caption{The two-point angular cross-correlation functions of SSA22 galaxies. The orange circles represent the cross-correlation function of LAEs and $z=3.1$ SMG candidates. Blue squares show the auto-correlation function of LAEs. These results indicate that the distributions of selected SMGs are correlated with those of $z=3.1$ LAEs.
For comparison, the cross-correlation functions of LAEs and the remaining 51 SMGs are also shown using cyan circles.
}
    \label{fig:ccf}
  \end{figure}  

\subsection{Cross-Correlation Function}

As mentioned above, the sky distribution of the $z=3.1$ candidate SMGs suggests
that a majority of SMGs are located in the most overdense region at the dozens-of-Mpc scale and that some SMGs lie in relatively less overdense regions. The two-point angular cross-correlation function is useful to evaluate statistically the degree of similarity in distributions between two types of galaxies, in this case SMGs and LAEs.
We calculate it in a way similar to the one taken in \cite{2009Natur.459...61T} and describe it only briefly here. The angular distances for all pairs between two populations are measured and the incremental quantity within a unit solid angle as a function of angular distance is calculated.
At this time, the estimator equation proposed by \cite{1993ApJ...412...64L} is utilized.
We confine the area so as not to generate artificial positive signals. Hereafter, we will consider the field observed by IRAC (See Figure 7) are considered.
Moreover, the pair which are separated by more than 15$^{\prime \prime}$ are excluded to avoid the source confusion effect.

The computed angular cross correlation function between 10 $z\sim3.1$ SMGs and 195 $z=3.1$ LAEs is shown in Figure \ref{fig:ccf}. 
For comparison, the auto-correlation functions of LAE is also shown. Error bars on the correlation functions were estimated from the root mean square among 5,000 bootstrap samples of the original catalogs.
Our results suggest that
$z\sim3.1$ SMGs are clustered and that their distribution is similar to that of $z=3.1$ LAEs, 
since we find positive amplitudes of the cross-correlation at angular distances $\leq$ 7 arcmin.
We also computed the cross-correlation function between the remaining 51 SMGs covered by IRAC and the $z=3.1$ LAEs similarly. In this case it is suggested that the distributions of the two are not correlated. 
The correlation between SMGs and LAEs in the SSA22 was firstly shown in \cite{2009Natur.459...61T}.
Though their selection of SMGs was based only on luminosity, they showed a possible correlation between the two samples.
Our work confirms the trend that they reported, constraining the photometric redshifts. Consequently, we arrive at a more sophisticated understanding of the correlation: the amplitude of the function is higher and the significance of the correlations is increased.

A similar approach using cosmological hydrodynamics simulations was followed by \cite{2012MNRAS.427.2866S}.
They also calculated the cross correlation function focusing on the $z=3.1$ universe in their simulation box, utilizing a code that successfully reproduced the observed source counts of SMGs.
 They indicated two trends. Firstly, they reported that there is a correlation between the distributions of SMGs and LAEs. Therefore, as a whole, our results are consistent with their simulations and suggest that SMGs selectively reside at relatively massive dark matter halos. The second trend is that the cross-correlation function between SMGs and LAEs has a profile similar to that of the correlation function between SMGs and dark matter halos, as presented in previous works (e.g., \citealt{2010Ap&SS.330..219S}).
The observed amplitudes of the cross correlation function of SMGs and LAEs are, however, much higher than that predicted from the simulation.
The reason is not apparent, but one possible interpretation is that in the SSA22 field the degree of mass assembly is much larger than what was assumed in the simulation, and as a consequence the degree of the clustering of galaxies becomes higher.

\section{Summary}

We imaged a 950 arcmin$^2$ field towards the SSA22 field covering a 1$\sigma$ depth down to $0.7-1.3$ mJy using the AzTEC 1.1mm camera attached on the ASTE.
This survey area corresponds to about 2.5 times of the previously reported area (390 arcmin$^2$) by \cite{2009Natur.459...61T}.
We detected 125 SMGs with S/N $\ge$ 3.5 and eight out of them are expected to be fake source arising from noise peaks.

We attempted to identify reliable counterparts to 125 AzTEC SMGs utilizing VLA 1.4 GHz imaging, MIPS 24 $\mu$m images and IRAC color diagnostics (\citealt{2008MNRAS.389..333Y}).
We considered the corrected Poissonian probability, $p$,
to evaluate the degree of chance coincidence.
We regarded counterpart candidates as robust counterparts if $p$ is less than 0.05.
Additionally sources with $0.05\leq p<0.20$ were considered as tentative counterparts.
We found that 59 SMGs have at least one reliable (i.e., robust or tentative) counterpart considering all diagnostics methods.

We performed SED fitting based on optical to near-infrared photometry utilizing the {\it HYPERZ} code, which provided us with photometric redshifts for 48 counterparts of 45 SMGs, which were all covered by all band IRAC observations.
We couldn't find any counterparts for 14 of 61 SMGs that had four-band IRAC coverage.
The redshift distribution of the SSA22 field is similar to those of GOODS-S AzTEC SMGs reported by \cite{2012MNRAS.420..957Y}.
These AzTEC SMGs tend to lie at a higher redshift universe than the radio-identified SCUBA SMGs ($z_{med}=2.2$ reported by \citealt{2005ApJ...622..772C}).
Some AzTEC sources without reliable counterparts may be located at higher-$z$ and would enhance this trend.
%Though the SSA22 field is known as a unique overdense region at $z=3.1$, we cannot find significant spike in the redshift distribution.

We found 10 AzTEC SMGs that possibly are at $z=3.1$ based on photometric and/or spectroscopic redshifts.
Among them, seven out of the 10 SMGs were concentrated into the core 12 Mpc $\times$ 12 Mpc region (comoving scale), which is consistent with the center of the galaxy distribution of the protocluster.
Cross-correlation functions indicate that the distribution of these 10 SMGs and that of $z=3.1$ LAEs are correlated.
These results suggest that SMGs are tend to be formed in extremely high density environments.

%and SMGs and LAEs co-exit within the large scale structure. 

\section*{Acknowledgments}
The ASTE project is driven by Nobeyama Radio Observatory (NRO), a branch of NAOJ, in collaboration with University of Chile, and Japanese institutes including University of Tokyo, Nagoya University, Osaka Prefecture University, Ibaraki University and Hokkaido University.
Observations with ASTE were in part carried out remotely from Japan by using NTT's GEMnet2 and its partner R\&E (Research and Education) networks, which are based on AccessNova collaboration of University of Chile, NTT Laboratories and NAOJ. AzTEC analysis performed at UMass is supported in part by NSF grant 0907952.
This study was supported
by the MEXT Grant-in-Aid for Specially Promoted Research (no.~20001003)
and the Grant-in-Aid for the Scientific
 Research from the Japan Society for the Promotion of Science
 (no.~19403005).
This paper is based on data collected at Subaru Telescope, which is operated by the National Astronomical Observatory of Japan. 
This work is based on observations made with the Spitzer Space Telescope, which is operated by the Jet Propulsion Laboratory, California Institute of Technology under a contract with NASA.
Some of the data reported here were obtained as part of the UKIRT Service Programme.
SI is financially supported by a Research Fellowship from the JSPS for Young Scientists. 
YT is supported by JSPS Grant-in-Aid for Research Activity Start-up (no. 23840007). 
KSS is supported by the National Radio Astronomy Observatory, which is a facility of the National Science Foundation operated under cooperative agreement by Associated Universities, Inc.

\bsp

\label{lastpage}

\end{document}